\def\draftversion{false}

\RequirePackage{ifthen}
\ifthenelse{\equal{\draftversion}{true}}{
  \documentclass[prb,galley,showpacs,preprintnumbers,citeautoscript,
      amsmath,amssymb]{revtex4}
}{
  \documentclass[citeautoscript,floatfix,aps,prb,twocolumn,
      superscriptaddress]{revtex4-1}
}

\usepackage{amsmath}
\usepackage{graphicx}
\usepackage{epstopdf}
\usepackage{natbib}
\usepackage{array}
\usepackage{dcolumn}
\usepackage{bm}

\usepackage{soul}  

\usepackage[usenames,dvipsnames]{color}

\soulregister\cite7

\ifthenelse{\equal{\draftversion}{true}}{
  \marginparwidth 2.7in
  \marginparsep 0.5in
  \newcounter{comm} 
  \def\commnext{\stepcounter{comm}}
  \def\commtext{{\bf\color{blue}[\arabic{comm}]}}
  \def\commmar{{\bf\color{blue}[\arabic{comm}]}}
  \def\dvm#1{\commnext\marginpar{\small DV\commmar: #1}\commtext}
  \def\cdm#1{\commnext\marginpar{\small CED\commmar: #1}\commtext}
  \def\msm#1{\commnext\marginpar{\small MS\commmar: #1}\commtext}
  \def\asm#1{\commnext\marginpar{\small AS\commmar: #1}\commtext}
  \def\mlab#1{\marginpar{\small\bf #1}}
  
}{
  \def\dvm#1{}
  \def\cdm#1{}
  \def\msm#1{}
  \def\asm#1{}
  \def\mlab#1{}
  
}

\begin{document}

\title{Metric-wave 
        approach to flexoelectricity within
        density-functional perturbation theory }

\author{Andrea Schiaffino}
\affiliation{Institut de Ci\`encia de Materials de Barcelona
(ICMAB-CSIC), Campus UAB, 08193 Bellaterra, Spain}
\author{Cyrus E. Dreyer}
\affiliation{Department of Physics and Astronomy, Rutgers University, Piscataway, New Jersey 08845-0849, USA}
\affiliation{Department of Physics and Astronomy, Stony Brook University, Stony Brook, New York 11794-3800,
USA}
\affiliation{Center for Computational Quantum Physics, Flatiron Institute, 162 5th Avenue, New York, NY 10010, USA}
\author{David Vanderbilt}
\affiliation{Department of Physics and Astronomy, Rutgers University, Piscataway, New Jersey 08845-0849, USA}

\author{Massimiliano Stengel}
\affiliation{ICREA - Instituci\'o Catalana de Recerca i Estudis Avan\c{c}ats, 08010 Barcelona, Spain}
\affiliation{Institut de Ci\`encia de Materials de Barcelona
(ICMAB-CSIC), Campus UAB, 08193 Bellaterra, Spain}

\date{\today}
\begin{abstract}

Within the framework of density functional perturbation theory (DFPT),
we implement and test a novel ``metric wave'' response-function approach.
It consists in the reformulation of an acoustic phonon perturbation in the curvilinear
frame that is comoving with the atoms. This means
that all the perturbation effects are encoded in the
first-order variation of the real-space metric,
while the atomic positions remain fixed.
This approach can be regarded as the generalization of the uniform
strain perturbation of Hamann \textit{et al.}
[D. R. Hamann, X. Wu, K. M. Rabe, and D. Vanderbilt,
  Phys. Rev. B \textbf{71}, 035117 (2005)]
to the case of inhomogeneous deformations,
and greatly facilitates the calculation of advanced
electromechanical couplings such as the flexoelectric tensor.
We demonstrate the accuracy of our approach with extensive tests
on model systems and on bulk crystals of Si and SrTiO$_3$.

\end{abstract}

\pacs{Valid PACS appear here}
\maketitle

  \section{Introduction}

  Theoretical attention to the fundamentals of
  mechanical deformations
  is of growing importance
  due to a surge of interest~\cite{Zubko07,Majdoub08,Narvaez16}
  in the flexoelectric effect, i.e., the polarization (${\bf P}$)
  response of a generic insulator to a strain gradient deformation.
  The renewed activity on flexoelectricity has mainly been motivated
  by a number of promising experimental results,
  demonstrating its large potential
  in different applications such as sensors and MEMS~\cite{Bhaskar2016},
  memory storage~\cite{Lu2012} and
  replacement of piezoelectrics~\cite{Cross06}.
  This experimental excitement motivates the
  urgency of supporting the results
  with a robust and predictive theory.

  In this context,
  density functional theory (DFT) appears as the most natural
  approach to study electromechanical response properties
  with unbiased quantum-mechanical
  accuracy.
  Techniques for calculating piezoelectricity (${\bf P}$
  response to uniform strain) are now well established~\cite{Martin1972,Hamann2005};
  however, generalization to flexoelectricity is far from trivial,
  and viable methodologies have started to appear only very
  recently \cite{Hong2013,Stengel2013,Stengel2014,Dreyer2018,quantum}.
  Their conceptual basis consists in the long wave analysis of
  acoustic phonon perturbations.
  This has the advantage of recasting a strain gradient,
  which breaks translational symmetry,
  into a periodic problem, by exploiting the standard
  treatment of incommentsurate perturbations
  within the context of density functional
  perturbation theory (DFPT).
  Since we are interested in the electric response
  of insulators to mechanical deformations, the
  relevant quantity on which we have to focus
  our attention is the polarization
  response.
  By taking its long-wave expansion,
  we can systematically identify the lowest orders in the wavevector
  ${\bf q}$ with specific electromechanical
  couplings~\cite{Stengel2013,Stengel2013natcom}.
  The zeroth-order term, related to rigid translations
  of the lattice, vanishes due to the acoustic sum rule;
  the first-order term is the piezoelectric coefficient;
  the second-order term corresponds to the
  flexoelectric coefficient.

  Generally speaking, the total polarization response of the crystal
  includes both purely electronic and lattice-mediated contributions.
  While the latter are relatively uncomplicated to understand
  and calculate, as the corresponding formulas bear many similarities to
  those that are valid for simplified point-charge models~\cite{Tagantsev-86},
  for the clamped-ion contribution to the polarization
  the knowledge of the microscopic current-density
  response~\cite{Stengel2013,Hong2013} to the deformation is needed.
  Despite the fundamental nature
  of this observable in quantum mechanics~\cite{Sakuri},
  the current density is not routinely available in public DFT codes.
  It is only recently that some of us established a computationally
  tractable
  definition of the current density
  and used it in the context of phonon perturbations
  to obtain the flexoelectric coefficients of selected
  materials.~\cite{Dreyer2018}
  This strategy represents a methodological breakthrough,
  as it allows for the first time the calculaton of all of the independent components of the
  bulk flexoelectric tensor by using a primitive crystal cell.

  There are further subtleties, however, that the approach of
  Ref.~\onlinecite{Dreyer2018} has addressed only partially.
For example, it has become clear in the past few years that flexoelectricity
is not a genuine bulk property:
before attempting any comparison between \emph{ab initio} results and experiments,
the bulk flexoelectric tensor
needs to be combined with the relevant surface contributions,
i.e., those coming from ``surface piezoelectricity.''~\cite{Stengel2013,Tagantsev12}
Surface effects might appear, at first sight, irrelevant in the context of
a bulk theory; yet
the separation between
surface and bulk contributions to the flexoelectric response is not unique.

  A manifestation of this issue was illustrated in
  Ref.~\onlinecite{quantum}, where it was shown that the electronic
  flexoelectric tensor consists of two distinct physical
  contributions: a ``static''  and a ``dynamic'' term.  The
  latter, in particular,
  is related to rotation gradients (a subset of the
  strain-gradient tensor components), and is proportional to the
  orbital magnetic susceptibility tensor.
  While both mechanisms contribute to the physical
  current-density field that is generated by a
  strain gradient, and therefore are implicitly
  present in the bulk response
  as calculated via a standard phonon perturbation,
  only the former is relevant to the electromechanical
  response of a finite sample. (The rotation-gradient
  contribution, being a purely solenoidal current,
  makes no contribution to the charge density in the \emph{bulk},
  while its effect at the boundary is
  exactly canceled by an equal and opposite
  \emph{surface} term.)
  Therefore, in Ref.~\onlinecite{Dreyer2018} an
  independent calculation of the diamagnetic susceptibility
  was performed in order to isolate the physically
  relevant static part.

  An alternative approach for calculating the electronic
  flexoelectric tensor consists in employing
  the time dependent Schr\"odinger equation rewritten
  in curvilinear coordinates~\cite{quantum}.
  These coordinates are identified by the frame
  that is co-moving with the atoms.
  In such a frame the atoms do not move by construction, and all the information
  on the perturbation is encoded in the macroscopic displacement
  field and its gradients (e.g., the metric tensor).
  For this reason, we identify the aforementioned representation
  of the acoustic phonon as a ``metric perturbation'';
  indeed, this constitutes
  a generalization of the metric tensor formulation of the
  uniform strain~\cite{Hamann2005}
  to a spatially modulated perturbation.
  The curvilinear frame is particularly convenient
  because it naturally separates the static and the dynamic
  contributions to the electronic flexoelectric
  tensor.
  Thus, one can readily use the metric perturbation
  to calculate the static contribution directly,
  and thereby eliminate the need for any post-processing step
  connected with the diamagnetic correction.
  Moreover, the metric perturbation is a computationally
  much more efficient approach to the calculation of the
  clamped-ion flexoelectric tensor, since
  it avoids the sum over individual sublattices which is
  implicit in the phonon approach of Ref.~\onlinecite{Dreyer2018}.
  A practical calculation of the flexoelectric tensor that takes
  full advantage of the curvilinear coordinates,
  however, has not yet been attempted.

  Part of the reason lies in some points of principle that were
  left unresolved in earlier works. First, the formalism of
  Ref.~\onlinecite{quantum} was derived under the assumption of an
  \emph{all-electron} description of the electronic structure, where
  the atoms are treated as point charges.
  This is clearly ill suited to
  a numerical implementation based on a plane-wave basis set. Prior
  to its practical use,
  the formalism needs to be generalized to the treatment of
  separable atomic pseudopotentials in the Kleinman-Bylander~\cite{Kleinman1982} form,
  at the very least.
  Second, the precise relationship between the first-order wave functions in
  the curvilinear and laboratory frames need to be established in order
  to firm up the conceptual foundations of the method. Based on earlier
  derivations,~\cite{quantum} for example, we know the relationship between
  the relevant physical observables (current density, flexoelectric tensor,
  and so forth).
  However, this was obtained in an idealized context of a complete basis set and
  continuous Brillouin-zone integration.
  This is not enough to predict
  how results will converge as a function of the common computational parameters,
  nor whether such convergence will be at all different compared to the phonon
  approach~\cite{Dreyer2018}.

  Here we address the above issues in full by deriving the missing
  pseudopotential terms in the metric-wave perturbation.
  Remarkably, we establish a rigorous link between the
  response in the Cartesian and co-moving frames, which shows that
  the respective first-order wave functions are related by
  a simple geometric contribution (i.e., one that can be
  expressed in terms of 
  ground-state quantities).
  We analyze the implications of this result for the
  observables of interest in the present context (charge density and
  current), leading naturally to a stringent numerical validation
  strategy.
  %

%
  Based on the aforementioned results, we then proceed to
  the code implementation and testing of the monochromatic metric perturbation,
  identified by a (generally) incommensurate wavevector ${\bf q}$, in the context of
  DFPT.
  The wave-function response to such a metric perturbation is then
  used as input for calculating the current-density response,
  as recently developed in Ref.~\onlinecite{Dreyer2018}.
  The resulting methodology allows flexoelectric coefficients
  to be calculated with unprecedented accuracy and computational efficiency.
  In particular, our numerical tests clearly demonstrate that the
  present method yields faster convergence with respect to
  ${\bf k}$-point mesh density and other computational parameters
  when compared with previous approaches.
  We rationalize this result in terms of the aforementioned
  relationship between the first-order wave functions in the
  curvilinear and laboratory frames.

  From the formal point of view, this work also establishes a
  direct link between the perturbative treatment of phonon and
  uniform strain perturbations, which previously have been regarded as two
  conceptually distinct sub-areas of DFPT.

  The paper is organized as follows.
  In Sec.~\ref{theo} we start by briefly motivating 
  the metric perturbation in the context of flexoelectricity.
  We then analyze in depth its connections with the established phonon
  and strain perturbations, while highlighting a number of important technical
  details related to the code implementation.
  In Sec.~\ref{tests} we present the results of our numerical tests, which
  we perform in the context of DFPT by
  calculating, among other properties, the flexoelectric tensor of selected
  representative crystals.
  Finally, in Sec.~\ref{con}, we present our summary and conclusions.
  \section{Theory\label{theo}}

 \subsection{Flexoelectric tensor\label{flexo}}

   The main motivation for the development and implementation of the metric
   perturbation comes from the practical calculation of the clamped-ion
   (i.e., purely electronic) flexoelectric tensor.
   Thus, to frame our arguments, in this subsection
   we shall start by briefly reviewing the existing~\cite{Stengel2013,Hong2013,Dreyer2018}
   theory of bulk flexoelectricity, and in particular, the phonon-based
   current-density approach of Ref.~\onlinecite{Dreyer2018}.
   Subsequently, we shall point out the advantages of the new metric
   perturbation, which we shall describe in further detail in Sec.~\ref{theory_metric}.

   The bulk flexoelectric (FxE) coefficients are given by
\begin{equation}
\label{muII}
\mu^{\text{I}}_{\alpha\beta,\omega\nu}=\frac{d
P_\alpha}{d\eta_{\beta,\omega\nu}},
\end{equation}
where $P_\alpha$ is the polarization in direction $\alpha$, and
\begin{equation}
\eta_{\beta,\omega\nu}=\frac{\partial^2u_\beta}{\partial
r_\omega\partial r_\nu}
\end{equation}
(i.e., defined as the second gradient of the displacement field $u_\beta$).
Alternatively, the FxE tensor can be written in type-II form as
\begin{equation}
\label{muIII}
\mu^{\text{II}}_{\alpha\beta,\omega\nu}=\frac{d
P_\alpha}{d\varepsilon_{\omega\nu,\beta}},
\end{equation}
where $\varepsilon_{\omega\nu,\beta}= \partial \varepsilon_{\omega\nu}/
\partial r_\beta$ is the first gradient of the symmetrized strain tensor ,
\begin{equation}
\varepsilon_{\omega\nu} = \frac{1}{2} \left( \frac{\partial u_\omega }{\partial r_\nu} +
                                             \frac{\partial u_\nu }{\partial r_\omega}  \right) =
                                             \frac{ h_{\omega\nu} + h_{\nu\omega}  }{2}.
\end{equation}
($h_{\omega\nu} =
  {\partial u_\omega }/{\partial r_\nu}$ is the unsymmetrized strain, also known as
  \emph{deformation gradient}). Choosing one or the other representation is a
  matter of convenience, as the independent entries of $\bm{\mu}^{\rm I}$ and
  $\bm{\mu}^{\rm II}$ are 
  linearly related to one another.
  Neither of the two
  definitions [Eq.~(\ref{muII}) or (\ref{muIII})], however, lends itself
  easily to a direct numerical implementation, as both involve an unbounded
  perturbation that breaks the translational symmetry of the lattice.

\subsubsection{Phonons\label{secPhon}}

To address this issue, Refs.~\onlinecite{Stengel2013,Hong2013} based their formalism on the cell average
of the microscopic polarization response to a monochromatic atomic distortion pattern,
\begin{equation}
 \overline{P}_{\alpha,\kappa \beta}^{\textbf{q}} = \frac{1}{\Omega} \int_{\rm cell} d^3r \, e^{-i{\bf q\cdot r} }
       \frac{\partial P_\alpha({\bf r})}{\partial \lambda_{\kappa\beta}},
\end{equation}
where the perturbation consists in a modulated displacement of the
sublattice $\kappa$ at the cell $l$ along the direction $\beta$ (${\bf R}_{l\kappa}$ indicates
the unperturbed lattice sites),
\begin{equation}
u^l_{\kappa \beta} = \lambda_{\kappa\beta} e^{ i{\bf q} \cdot {\bf R}_{l\kappa} }.
\label{perturb}
\end{equation}
The clamped-ion type-I flexoelectric coefficients can then be written as the
second gradient with respect to the wavevector ${\bf q}$ of the
aforementioned polarization response~\cite{Stengel2013,Hong2013},
\begin{equation}
\label{muI}
\begin{split}
\mu^{\text{I}}_{\alpha\beta,\omega\nu}&= -\frac{1}{2} \sum_\kappa \frac{\partial^2\overline{P}_{\alpha,\kappa \beta}^{\textbf{q}}}{\partial
q_\omega\partial q_\nu}.
\end{split}
\end{equation}
(Throughout this Section we shall consistently use the type-I 
representation since it is the most convenient for performing
the calculations, and we shall drop
the superscript for conciseness.
However, when presenting our results in Section~\ref{flexo_results},
we shall switch to type-II form, following the conventions of earlier works.)
Based on the current-density implementation of
Ref.~\onlinecite{Dreyer2018}, one can write
$\overline{P}_{\alpha,\kappa \beta}^{\textbf{q}}$ as a second-order
matrix element,
\begin{equation}
\begin{split}
\label{PqTR0}
  \overline{P}_{\alpha,\kappa \beta}^{\textbf{q}}
&=\frac{4}{N_k}\sum_{n\textbf{k}}  \langle u_{n\textbf{k}}\vert\hat{\mathcal{J}}_\alpha^{\textbf{k},\textbf{q}}
\vert\delta u_{n\textbf{k},\textbf{q}}^{\tau_{\kappa\beta}}\rangle,
\end{split}
\end{equation}
involving the ground-state Bloch orbitals, $|u_{n\textbf{k}}\rangle$, the current-density operator,
$\hat{\mathcal{J}}_\alpha^{\textbf{k},\textbf{q}}$, and the \emph{adiabatic}~\cite{Dreyer2018} wavefunction
response to the perturbation of Eq.~(\ref{perturb}),
$\vert\delta u_{n\textbf{k},\textbf{q}}^{\tau_{\kappa\beta}}\rangle$.
(The latter is defined, in the context of adiabatic perturbation theory,
as the change in the wavefunction to first order in the rate of change
of the phonon mode amplitude.)
In Ref.~\onlinecite{Dreyer2018}, some of us have implemented
and tested Eq.~(\ref{PqTR0}), and used it to calculate the flexoelectric
coefficients via Eq.~(\ref{muI});
we refer the interested reader to that work for the technical details.

\subsubsection{Rotation gradients\label{secRot}}

Many types of strain gradient involve a spatial variation in the local
\emph{rotation} as described by
the antisymmetric part of the
deformation gradient, $R_{\omega \nu} =
\left( {\partial u_\omega }/{\partial r_\nu} -
{\partial u_\nu }/{\partial r_\omega}  \right)/2$.
It has recently become clear~\cite{quantum,Dreyer2018} that rotation gradients produce,
in addition to other effects, a divergenceless current-density field, which contributes to the
bulk flexoelectric tensor proportionally to the diamagnetic susceptibility of the material.~\cite{quantum}
These contributions are implicitly present in Eq.~(\ref{muI});
however, since they ultimately will cancel
out with an equal and opposite surface term that originates from the
same physical mechanism, it is best to subtract them once and for
all. We can do this by defining the effective (type-I) flexoelectric
tensor
\begin{equation}
\label{pchi1}
\begin{split}
\mu'_{\alpha\beta,\omega\nu}
&= \mu_{\alpha\beta,\omega\nu} - \frac{1}{2}
\sum_{\gamma\lambda}\left(\epsilon^{\alpha\omega\gamma}\epsilon^{\beta\lambda\nu}+\epsilon^{\alpha\nu\gamma}\epsilon^{\beta\lambda\omega}\right)
 \chi^{\rm mag}_{\gamma\lambda}
\end{split}
\end{equation}
where $\epsilon$ is the Levi-Civita symbol and
$\chi^{\rm mag}_{\gamma\lambda}=\partial M_\gamma/\partial H_\lambda$
is the magnetic susceptibility tensor
(${\bf M}$ is the magnetization and ${\bf H}$ is the magnetic field).
This was the approach taken
in Ref.~\onlinecite{Dreyer2018}, where an independent calculation of $\chi^{\rm mag}_{\gamma\lambda}$
was performed in order to the obtain $\mu'_{\alpha\beta,\omega\nu}$.

\subsubsection{Metric response\label{metres}}

The procedure described in Secs.~\ref{secPhon} and
\ref{secRot} presents two drawbacks. First, an individual phonon
response calculation needs to be performed for each sublattice
$\kappa$. Second, an additional calculation of
$\chi^{\rm mag}_{\gamma\lambda}$ needs to be performed.
Both drawbacks were 
resolved in Ref.~\onlinecite{quantum} by
establishing an alternative formulation of the polarization response to an acoustic phonon,
\begin{equation}
\begin{split}
\label{PqTR}
  \overline{P}_{\alpha,\beta}^{\,\prime\,\textbf{q}}
&=\frac{4}{N_k}\sum_{n\textbf{k}}  \langle u_{n\textbf{k}}\vert\hat{\mathcal{J}}_\alpha^{\textbf{k},\textbf{q}}
\vert\delta u_{n\textbf{k},\textbf{q}}^{(\beta)}\rangle.
\end{split}
\end{equation}
   Here, in contrast to Eq.~(\ref{PqTR0}),
   the ket is the adiabatic
   response to a metric wave, i.e., an acoustic phonon
   perturbation described in the co-moving frame. Thus, the contributions
   from the individual atomic sublattices have implicitly been summed over.
Furthermore, the resulting
polarization is already free from the rotation-gradient contribution
described in the previous paragraph, and therefore we can directly write
\begin{equation}
\label{muIprime}
\begin{split}
\mu'_{\alpha\beta,\omega\nu}&= -\frac{1}{2}  \frac{\partial^2\overline{P}_{\alpha,\beta}^{\,\prime\,\textbf{q}}}{\partial
q_\omega\partial q_\nu},
\end{split}
\end{equation}
eliminating the need for an explicit calculation of $\chi^{\rm mag}_{\gamma\lambda}$.
Note that the metric formalism introduces another technical simplification at the level
of the current-density operator.
In the phonon case, an expression for
$\hat{\mathcal{J}}_\alpha^{\textbf{k},\textbf{q}}$ that is correct
up to second order in \textbf{q} was needed.\cite{Dreyer2018} Here
by contrast, since the wave-function response vanishes at
${\bf q}=0$, $\hat{\mathcal{J}}_\alpha^{\textbf{k},\textbf{q}}$ is only needed
\emph{up to first order},
\begin{equation}
\begin{split}
\label{ICLimp}
\hat{\mathcal{J}}_\alpha^{\textbf{k},\textbf{q}}&=-\Bigg(\hat{p}_\alpha^{\textbf{k}}+\frac{q_\alpha}{2}+\frac{\partial \hat{V}^{\textbf{k}}_{\text{NL}}}{\partial k_\alpha}+\frac{1}{2}\sum_{\gamma=1}^3 q_\gamma\frac{\partial^2 \hat{V}^{\textbf{k}}_{\text{NL}}}{\partial k_\alpha\partial k_\gamma} \Bigg),
\end{split}
\end{equation}
where $\hat{p}_\alpha^{\textbf{k}}=-i\hat{\nabla}_\alpha+k_\alpha$ is the cell-periodic momentum operator and $\hat{V}^{\textbf{k}}_{\text{NL}}=e^{i\textbf{k}\cdot\hat{\textbf{r}}}\hat{V}_{\text{NL}}e^{-i\textbf{k}\cdot\hat{\textbf{r}}}$, where $\hat{V}_{\text{NL}}$  nonlocal external potential operator.
In this work, we will use the ``straight-line path'' form of the current operator\cite{ICL2001,Essin2010}, although
such a choice is irrelevant for materials with cubic symmetry~\cite{Dreyer2018}.

In the following we shall delve into the technical details of the metric
perturbation within the general framework of DFPT.

\begin{figure*}
   \begin{center}
   \begingroup
    \setlength{\tabcolsep}{15pt} 
   \begin{tabular}{ccc}
   \includegraphics[width=1.8in]{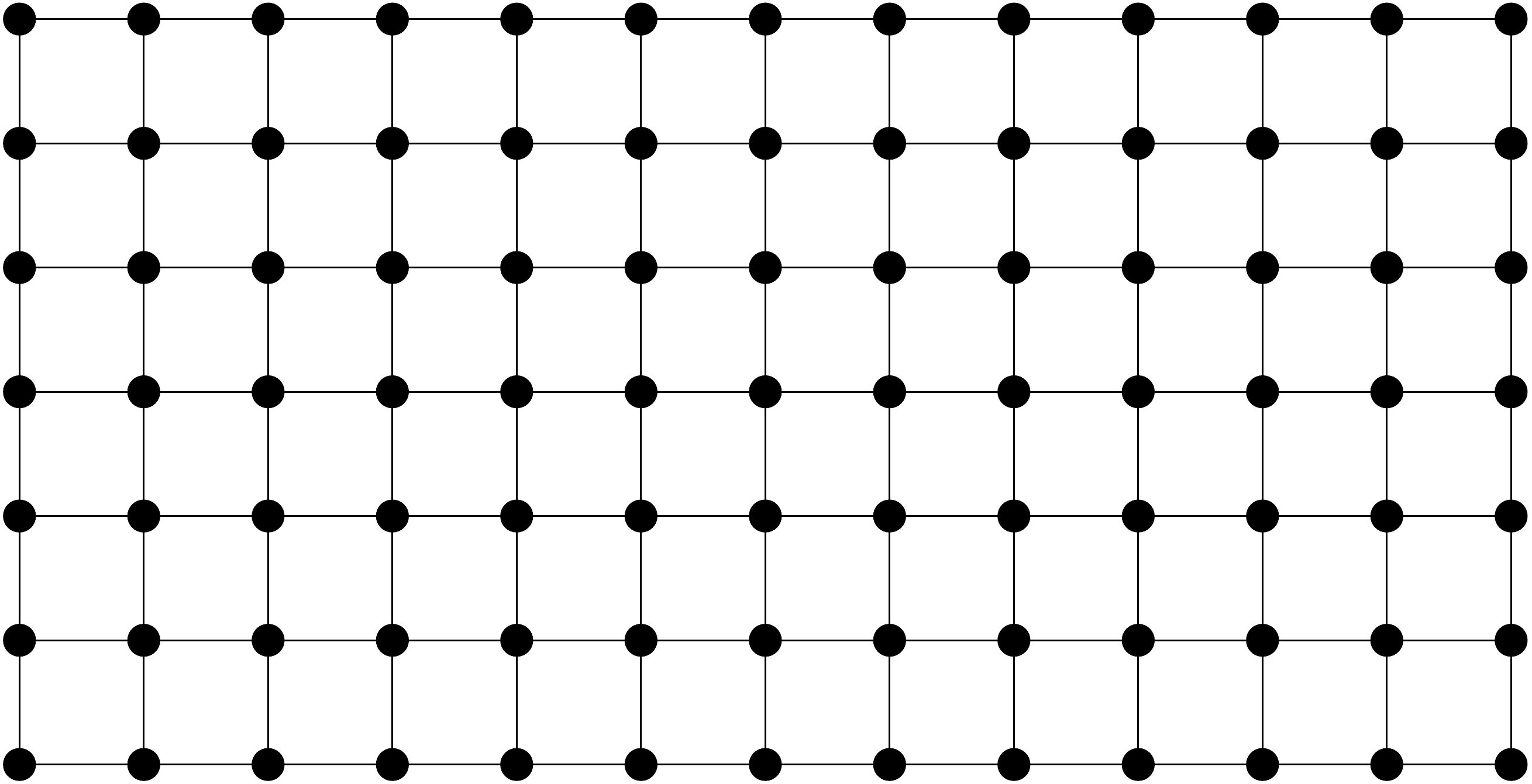} &
   \includegraphics[width=1.8in]{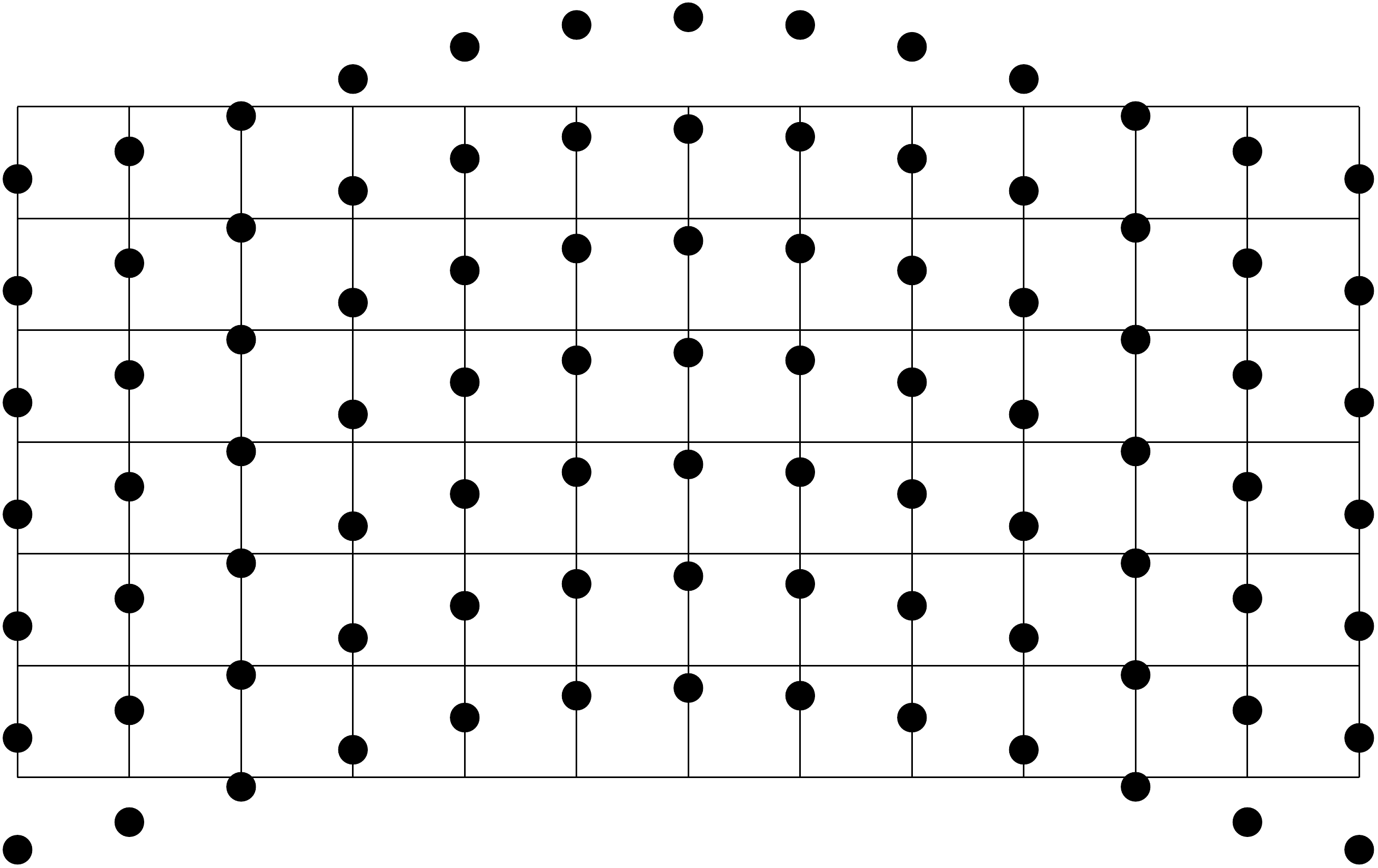} &
   \includegraphics[width=1.8in]{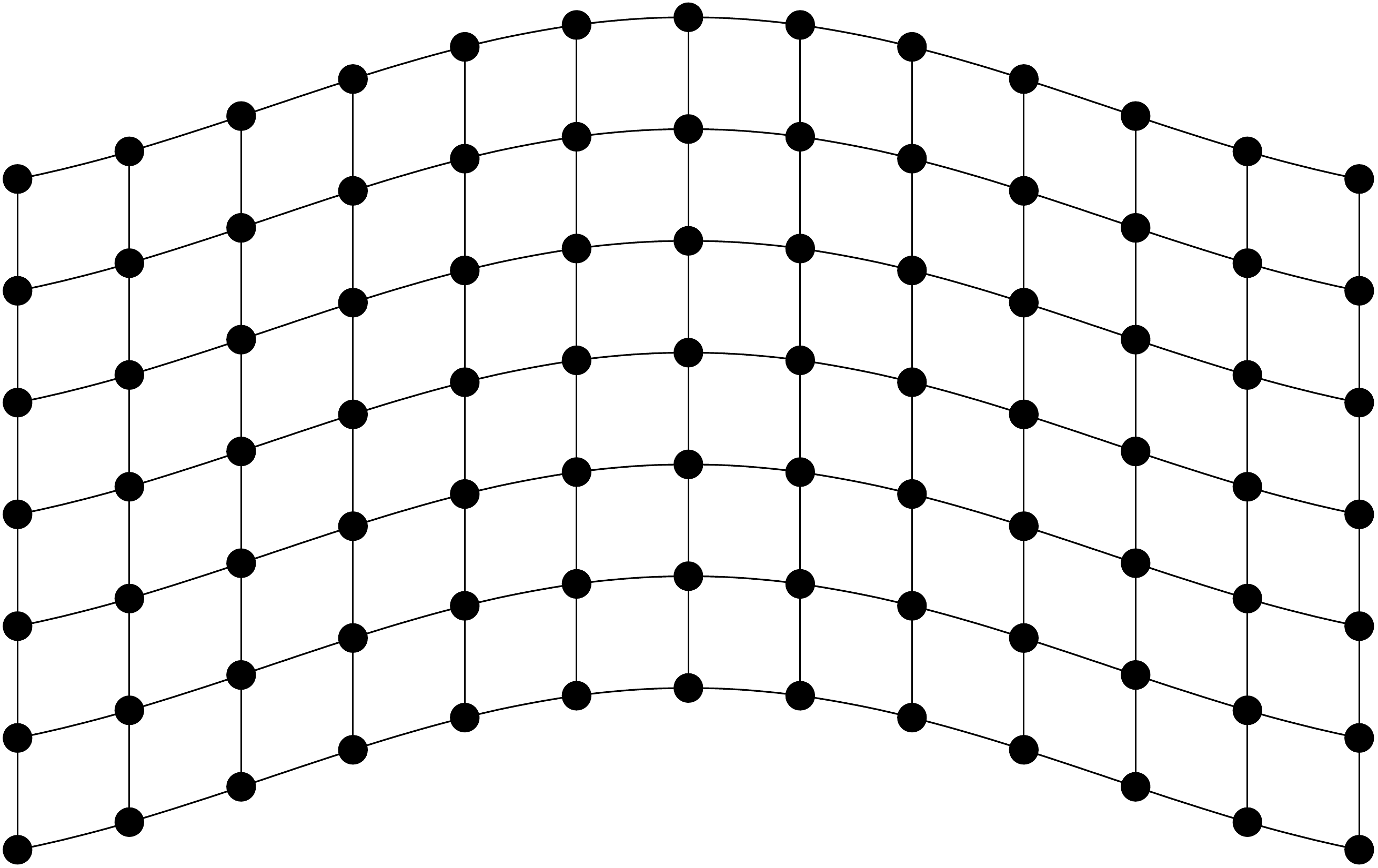} \\
    (a) & (b) & (c)
   \end{tabular}
   \endgroup
   \end{center}
   \caption{\label{fig:transform}
             Illustration of the coordinate transformation to the co-moving frame.
             (a) Unperturbed crystal lattice; black circles represent the atomic sites,
             horizontal and vertical lines represent the coordinate system. (b) Transverse
             acoustic phonon in the laboratory frame. (c) The same phonon in the curvilinear frame;
             note that the atoms do not move in this coordinate system -- the mechanical deformation is
             described via the metric.
             }
   \end{figure*}

  \subsection{Metric perturbation\label{theory_metric}}

  The starting ingredient for the metric perturbation is a ``clamped-ion'' acoustic
  phonon. This is a collective lattice mode where all atoms are perturbed according
  to Eq.~(\ref{perturb}) by using the same displacement amplitude for all sublattices,
  $\lambda_{\kappa \beta} = \lambda_\beta$.
  Next, we describe such a monochromatic acoustic wave \emph{in the curvilinear frame
  that is co-moving with the atoms} (Fig.~\ref{fig:transform}).
  This means that we combine the aforementioned displacement pattern with
  a simultaneous coordinate transformation that brings every atom
  back to its original position,
  \begin{equation}
  \label{coord}
  x_\beta(\bm{\xi}) = \xi_\beta + \lambda_\beta  e^{i \bm{\xi} \cdot {\bf q} }.
  \end{equation}
It is easy to verify that in the curvilinear reference, spanned by $\bm{\xi}$,
the atoms do not move; the perturbation now concerns the metric of the deformation.

  The unperturbed Kohn-Sham Hamiltonian reads as
  \begin{equation}
  \label{Hunpert}
  \hat{H}^{(0)} = \hat{T}^{(0)}  + \hat{V}^{\text{Hxc},(0)}+  \hat{V}^{\text{psp},(0)},
  \end{equation}
  where $\hat{T}^{(0)}$, $\hat{V}^{\text{Hxc},(0)}$ and $\hat{V}^{\text{psp},(0)}$ are the kinetic,
  exchange-correlation and pseudopotential terms, respectively.
  The latter consists in a local and a separable contribution,
  \begin{equation}
  \label{Vlocsep}
  V^{\text{psp},(0)}({\bf r,r'}) = V^{\text{loc},(0)}({\bf r}) \delta({\bf r-r'}) +
                                   V^{\text{sep},(0)}({\bf r,r'}),
  \end{equation}
  both written as lattice sums of individual atomic contributions, e.g.,
  \begin{equation}\label{Vloccart}
  V^{\text{loc},(0)}({\bf r}) = \sum_{l\kappa} v^{\text{loc}}_\kappa ({\bf r}-{\bf R}_{l\kappa}),
  \end{equation}
  where ${\bf R}_{l\kappa} = {\bf R}_{l} +\bm{\tau}_\kappa$.
The separable pseudopotential term is written in the Kleinman-Bylander (KB)
 form,
 \begin{equation}
   V^{\text{sep},(0)}({\bf r},{\bf r}')=\sum_{l\kappa\mu} e_{\mu \kappa}\zeta_{\mu \kappa}({\bf r} - {\bf R}_{l\kappa})\zeta_{\mu\kappa}^*({\bf r}'- {\bf R}_{l\kappa}),
 \end{equation}
 where $\zeta_{\mu \kappa}({\bf r})$ are the KB projectors, indexed by $\mu$,
 and $e_{\mu \kappa}$ are the corresponding coefficients.

 Subsequently to the change of coordinates,
 we shall write the ``static'' first-order Hamiltonian
 in the curvilinear frame as
 \begin{equation}
 \mathcal{\hat{H}}^{(\beta)}_{\bf k,q} = \hat{H}^{(\beta)}_{\bf k,q} + \hat{V}^{(\beta)}_{\bf q},
 \end{equation}
 i.e., as the sum of an ``external potential'' $\hat{H}^{(\beta)}_{\bf k,q}$
 plus a self-consistent contribution,
\begin{equation}
   V^{(\beta)}_{\bf q} ({\bf r}) = \int d^3 r' \, K_{\rm Hxc}({\bf r,r'}) e^{i{\bf q} \cdot ({\bf r'-r})} n^{(\beta)}_{\bf q}({\bf r'}),
\end{equation}
 that depends on the first-order charge density,
\begin{equation}
 n^{(\beta)}_{\bf q}({\bf r}) = \frac{4}{N_{\bf k}}  \sum_{m\bf k} \langle u^{(0)}_{m \bf k}| {\bf r} \rangle \langle {\bf r} |  u^{(\beta)}_{m\bf k,q} \rangle,
 \end{equation}
 via the Hartree, and exchange-correlation kernel,
 \begin{equation}
K_{\rm Hxc}({\bf r,r'}) =  \frac{\delta V_{\rm Hxc}({\bf r})}{\delta n({\bf r}')} \Big|_{n^{(0)}} =
 \frac{\delta^2 E_{\rm Hxc}}{\delta n({\bf r}) \delta n({\bf r}')} \Big|_{n^{(0)}}.
\end{equation}
In contrast to
most perturbations, however, the external potential
here takes contributions from all individual pieces of the Hamiltonian,
including the kinetic, pseudopotential, Hartree and exchange-correlation terms.
(The situation is analogous to the strain perturbation introduced by
Hamann {\em et al.},\cite{Hamann2005} for which the same considerations hold.)
In particular, loosely following Ref.~\onlinecite{quantum}, we shall write
\begin{equation}
\label{Hbeta}
\hat{H}^{(\beta)}_{\bf k,q} =
\hat{T}^{(\beta)}_{\bf k,q} + \hat{V}_{\bf k,q}^{{\rm psp},(\beta)}
 + \hat{V}_{\bf q}^{{\rm H0},(\beta)} + \hat{V}_{\bf q}^{{\rm XC0},(\beta)} + \hat{V}_{\bf q}^{{\rm geom},(\beta)} .
\end{equation}
In Eq.~(\ref{Hbeta}),
\begin{equation}
\hat{T}_{\bf k,q}^{(\beta)} = -\frac{i}{2} \left[  (\hat{p}_{{\bf k}\beta} +  q_\beta   ) \, {\bf q} \cdot  \hat{\bf p}_{{\bf k}}     +
                                                                     (\hat{\bf p}_{{\bf k}} +  {\bf q}   ) \cdot  {\bf q} \, \hat{p}_{{\bf k}\beta}   \right],
\end{equation}
is the kinetic term
($\hat{p}_{{\bf k}\beta} = -i \partial / \partial \xi_\beta + k_\beta$ is the
 canonical momentum operator in curvilinear space).
 For notational purposes we shall write the
 remainder of the contributions as matrix elements on two plane waves, e.g.,
 \begin{equation}
 W_{\bf k,q}^{(\beta)}({\bf G, G}') = \langle {\bf G+k+q}| \hat{W}_{\bf k,q}^{(\beta)} | {\bf G'+k} \rangle
 \end{equation}
 for an arbitrary operator $\hat{W}$.

 Regarding the pseudopotential term, we operate the same decomposition as
 in Eq.~(\ref{Vlocsep}),
\begin{equation}
V_{\bf k,q}^{{\rm psp},(\beta)}({\bf G, G}') = V_{\bf q}^{{\rm loc},(\beta)}({\bf G-G}') + V_{\bf k, q}^{{\rm sep},(\beta)}({\bf G,G}')
\end{equation}
i.e. we write it as the sum of a local
\begin{equation}\label{Vloc}
\begin{split}
V_{\bf q}^{{\rm loc},(\beta)}( {\bf G} ) &=
 i G_\beta \frac{1}{\Omega} \sum_\kappa e^{-i {\bf G} \cdot \bm{\tau}_\kappa } v^{\rm loc}_{\kappa} ({\bf G})  \\
& -i (G_\beta + q_\beta) \frac{1}{\Omega} \sum_\kappa e^{-i {\bf G} \cdot \bm{\tau}_\kappa } v^{\rm loc}_{\kappa} ({\bf G + q}),
\end{split}
\end{equation}
and a nonlocal contribution,
\begin{widetext}
\begin{equation}\label{Vsep}
\begin{split}
V_{{\bf k,q}}^{{\rm sep},(\beta)} ({\bf G,G}') = \frac{1}{\Omega}
 \sum_{\mu \kappa} e_{\mu \kappa} e^{-i({\bf G - G}') \cdot \bm{\tau}_\kappa }
       \Big\{ & i(G_\beta + k_\beta + \frac{q_\beta}{2}) \zeta_{\mu \kappa} ({\bf G+k})\zeta^*_{\mu \kappa} ({\bf G'+k})  \\
 & -  i(G_\beta - G'_\beta + q_\beta) \zeta_{\mu \kappa} ({\bf G+k+q})\zeta^*_{\mu \kappa} ({\bf G'+k}) \\
 &   - i(G'_\beta + k_\beta + \frac{q_\beta}{2}) \zeta_{\mu \kappa} ({\bf G+k+q})\zeta^*_{\mu \kappa} ({\bf G'+k+q}) \Big\},
\end{split}
\end{equation}
\end{widetext}
where $\zeta_{\mu \kappa} ({\bf G+k})$ indicates the Fourier components of the KB projectors.

The two terms
\begin{eqnarray}
 V_{\bf q}^{{\rm H0},(\beta)}({\bf G}) &=&   4\pi i \left( -\frac{G_\beta + q_\beta}{|{\bf G+q}|^2} + \frac{G_\beta}{G^2} \right) n^{(0)}({\bf G}), \label{VH0}      \\
 V_{\bf q}^{{\rm XC0},(\beta)}({\bf G}) &= &  -i q_\beta V^{\rm xc,(0)}({\bf G})
\end{eqnarray}
are the ``geometric'' (i.e., only depending on the unperturbed quantity, $n^{(0)}$)
contributions to the
Hartree (H) and exchange-correlation (XC) potentials, respectively.
Finally,
\begin{equation}
V^{{\rm geom},(\beta)}_{\bf q} = -\frac{i}{4} q_\beta \, q^2
\end{equation}
is an additional geometric potential
originating from the change of coordinates, which we introduce
here for completeness (this structureless potential is irrelevant for either the uniform strain
or the strain-gradient response, as it is of third order in ${\bf q}$).

Explicit derivations of most of the above expressions can be found in
Ref.~\onlinecite{quantum}. Regarding the
pseudopotential pieces, which have been derived here, some
additional details can be found in Appendix \ref{app:pseudo}.

Based on the above, it is now easy to demonstrate the following points:
\begin{itemize}
\item For an arbitrary ${\bf q}$, $V^{\rm H0,(\beta)}_{\bf q}({\bf G})$ exactly matches the
       metric contribution to the electrostatic potential as derived in
       Ref.~\onlinecite{quantum}. 
\item The external perturbation $\hat{H}^{(\beta)}_{\bf k,q}$ identically vanishes in the limit ${\bf q}=0$.
\item The first ${\bf q}$-gradient of the above expressions recovers the HWRV~\cite{Hamann2005} treatment
      of the uniform strain,
\begin{equation}\label{Hmetricstain}
 \hat{\mathcal{H}}^{(\beta)}_{\bf k,\gamma} = i \hat{\mathcal{H}}^{(\beta \gamma)}_{\bf k},
\end{equation}
and is symmetric under $\beta \gamma$ exchange.

\end{itemize}
We have, therefore, achieved the
desired generalization of the HWRV metric tensor formalism to a monochromatic
displacement wave of arbitrary ${\bf q}$.

  \subsection{Relationship to the response in the laboratory frame}\label{subsec:rel_two_frames}

In this section we shall establish the explicit
link between the metric perturbation described in the previous subection
(which, as we said, is defined in the co-moving frame and reduces to a uniform
strain perturbation in the long-wave limit) and the familiar phonon perturbation,
which is defined in the laboratory frame.
In particular, we shall show that the corresponding response
functions (``metric'' versus ``phonon'') differ by a geometric
piece that depends on the ground-state orbitals only.
These analytical results will prove to be important for
testing our numerical implementation, as we shall see shortly.
They also provide an interesting
formal unification of two areas of DFPT (related to the response to
phonons and strains, respectively) that were formerly regarded as
conceptually distinct.

The first-order external potential for a 
phonon perturbation
consists of a local potential plus a separable contribution,
\begin{equation}
H^{\tau_{\kappa \beta}}_{\bf k,q} ({\bf G,G'})
   = V_{\bf q}^{{\rm loc},\tau_{\kappa \beta}}({\bf G-G'}) + V_{\bf k,q}^{{\rm sep},\tau_{\kappa\beta} }({\bf G,G'}),
\label{htau}
\end{equation}
where
\begin{widetext}
\begin{eqnarray}
{V_{\bf q}^{{\rm loc},\tau_{\kappa \beta} }}({\bf G}) &=& -i (G_\beta + q_\beta) \frac{1}{\Omega}  
e^{-i{\bf G} \cdot \bm{\tau}_\kappa} \, v^{\rm loc}_\kappa ({\bf G + q}), \label{vloc} \\
V^{{\rm sep},\tau_{\kappa \beta}}_{\bf k,q}({\bf G},{\bf G}') &=& -i(G_\beta + q_\beta - G'_\beta)
\frac{1}{\Omega} \sum_{\mu} e^{-i({\bf G-G'})\cdot \bm{\tau}_\kappa} \,
e_{\mu \kappa} \zeta_{\mu \kappa}({\bf k +q + G}) \zeta^*_{\mu \kappa}({\bf k + G}').
\label{vsep}
\end{eqnarray}
\end{widetext}
  Note that in Eqs.~(\ref{vloc}) and (\ref{vsep}), the structure
  factors differ slightly from those that are commonly implemented in
  DFPT (e.g., Ref.~\onlinecite{Gonze1997}), which read as
  $e^{-i({\bf G+q}) \cdot \bm{\tau}_\kappa}$ and
  $e^{-i({\bf G-G'+q})\cdot \bm{\tau}_\kappa}$, respectively. This
  difference is a consequence of the fact that here we have introduced
  an extra phase, $e^{i{\bf q\cdot \tau}_{\kappa}}$, into the
  monochromatic phonon perturbation~\cite{Stengel2013}.

In the
laboratory frame, an acoustic phonon perturbation can be readily
constructed as a sublattice sum of the above,
\begin{equation}
H^{u_{\beta}}_{\bf k,q}({\bf G},{\bf G}') = \sum_\kappa H^{\tau_{\kappa \beta}}_{\bf k,q}({\bf G},{\bf G}').
\end{equation}
Here and in the following we use the symbol $u_\beta$ to indicate a
laboratory-frame acoustic phonon perturbation, not to be confused with the
corresponding metric perturbation labeled by $(\beta)$.
The corresponding first-order wavefunctions satisfy the following
Sternheimer equation~\cite{Gonze1997},
\begin{equation}
\left( \hat{H}_{\bf k+q}^{(0)} + a\hat{P}_{\bf k+q} - \epsilon^{(0)}_{m\bf k} \right) | u^{u_\beta}_{m {\bf k,q}} \rangle =
  -\hat{Q}_{\bf k+q} \hat{\mathcal{H}}^{u_{\beta}}_{\bf k,q} | u^{(0)}_{m {\bf k}} \rangle,
  \label{sternheimer}
\end{equation}
where $\hat{\mathcal{H}}^{u_{\beta}}_{\bf k,q}$ is, as usual, the
self-consistent counterpart of $\hat{H}^{u_{\beta}}_{\bf k,q}$.  Note
that an acoustic phonon physically reduces to a rigid translation of
the whole crystal lattice at $\mathcal{O}(q^0)$. At $\mathcal{O}(q^1)$
and $\mathcal{O}(q^2)$, respectively, it should provide complete
information about the response to a uniform strain and strain gradient
deformation.

To see the relationship between the laboratory and curvilinear frame
pictures, it is convenient to take one step back, and consider the
first-order Hamiltonians in the original Hilbert space, i.e., without
factoring out the incommensurate phases that belong either to the
Bloch orbitals or to the first-order Hamiltonian.
(Recall that the first-order Hamiltonian $\hat{\mathcal{H}}^{\lambda}$
is related to its periodic part in momentum space as
$\hat{\mathcal{H}}^{\lambda}_{\bf k,q} = e^{-i({\bf k+q})\cdot{\bf r}}
\, \hat{\mathcal{H}}^{\lambda} \, e^{i{\bf k}\cdot{\bf r}}$.)  We
shall postulate (and later prove that it is consistent with the
results derived so far) the relationship
\begin{equation}
\hat{\mathcal{H}}^{u_{\beta}} = \hat{\mathcal{H}}^{(\beta)}
 + i \left[\hat{H}^{(0)}, \frac{1}{2} \left( e^{i{\bf q}\cdot{\bf r}} \hat{p}_\beta + \hat{p}_\beta e^{i{\bf q}\cdot{\bf r}}  \right) \right].
\label{postulate}
\end{equation}
One can recognize in the commutator the gauge-field contribution to the
perturbation in curvilinear coordinates discussed in Ref.~\onlinecite{quantum},
\begin{equation}
\hat{\mathcal{H}}^{\dot{\lambda}_\beta}({\bf q})
  = -\frac{1}{2} \left( e^{i{\bf q}\cdot{\bf r}} \hat{p}_\beta + \hat{p}_\beta e^{i{\bf q}\cdot{\bf r}}  \right).
\end{equation}
[By taking the momentum-space representation of the above operator, $\hat{\mathcal{H}}^{\dot{\lambda}_\beta}_{\bf k,q} =
e^{-i{\bf (k+q) \cdot r}} \hat{\mathcal{H}}^{\dot{\lambda}_\beta}({\bf q}) e^{i{\bf k\cdot r}}$, one readily recovers
Eq.~(90) of Ref.~\onlinecite{quantum}.]
This clarifies the
physical interpretation of Eq.~(\ref{postulate}) as being closely linked to
the coordinate change discussed in Ref.~(\onlinecite{quantum}).
Then, after reverting to our previous notation,
the relation between laboratory-frame
and metric responses in Eq.~(\ref{postulate}) becomes
\begin{equation}
\hat{\mathcal{H}}^{u_{\beta}}_{\bf k,q} =
  \hat{\mathcal{H}}^{(\beta)}_{\bf k,q}  + i \hat{H}^{(0)}_{\bf k + q} \left( \hat{p}_{\bf k\beta} + \frac{q_\beta}{2} \right)
  -i \left( \hat{p}_{\bf k\beta} + \frac{q_\beta}{2} \right) \hat{H}^{(0)}_{\bf k}.
\label{transform}
\end{equation}
The correctness of this result can be verified by comparing the
explicit formulas for the perturbed Hamiltonians piece by piece.
In particular, the second and the third terms on the rhs of Eq.~(\ref{transform})
 precisely cancel the kinetic and geometric contributions in $\hat{\mathcal{H}}^{(\beta)}_{\bf k,q}$,
and they also account for the difference between the pseudopotential,
Hartree, and XC terms in $\hat{\mathcal{H}}^{(\beta)}_{\bf k,q}$ and
$\hat{\mathcal{H}}^{u_{\beta}}_{\bf k,q}$.

If we now plug Eq.~(\ref{transform}) into Eq.~(\ref{sternheimer}), we obtain
an analogous Sternheimer equation with $\hat{\mathcal{H}}^{(\beta)}_{\bf k,q}$
replacing $\hat{\mathcal{H}}^{u_{\beta}}_{\bf k,q}$, and
with the laboratory-frame first-order wavefunctions
related to the metric ones by
\begin{eqnarray}
|u^{u_\beta}_{m {\bf k,q}} \rangle &=& |u^{(\beta)}_{m {\bf k,q}} \rangle  + |\Delta u^\beta_{m {\bf k}} \rangle,
\label{umet}
\end{eqnarray}
where $|\Delta u^\beta_{m {\bf k}} \rangle$ is a purely geometric
(i.e., defined in terms of the ground-state orbitals only) contribution,
\begin{eqnarray}
 |\Delta u^\beta_{m {\bf k}} \rangle &=& -i \hat{Q}_{\bf k+q}\left( \hat{p}_{\bf k\beta} + \frac{q_\beta}{2} \right) | u^{(0)}_{m {\bf k}} \rangle.
  \label{deltau}
\end{eqnarray}
This constitutes the main result of this Section.

To illustrate its physical meaning
it is useful, first of all, to calculate the contribution of $|\Delta u^{\beta}_{m {\bf k,q}} \rangle$
to the first-order electron density, and check whether it matches our expectations for the
relationship between its laboratory-frame and curvilinear-frame representations.
To this end, recall the definition of the density response to a generic perturbation
$\lambda$,
\begin{equation}
n^{\lambda}_{\bf q}({\bf r}) = \frac{4}{N_{\bf k}} \sum_{n\bf k} \langle u^{(0)}_{n\bf k} | {\bf r} \rangle
   \langle {\bf r} | u^{\lambda}_{n\bf k,q} \rangle.
\label{nlam}
\end{equation}
By combining Eq.~(\ref{nlam}) with Eq.~(\ref{umet}), we find
\begin{eqnarray}\label{rhorelation}
n^{u^{\beta}}_{\bf q}({\bf r}) &=& n^{(\beta)}_{\bf q}({\bf r}) + \Delta n^{\beta}_{\bf q}({\bf r}),
\end{eqnarray}
where $\Delta n^{\beta}_{\bf q}({\bf r})$ is, again, a purely geometric
object. One can arrive at an explicit formula after observing that
$\hat{Q}_{\bf k+q} = 1 - \hat{P}_{\bf k+q}$; this leads to two
separate contributions to $\Delta n^{\beta}_{\bf q}({\bf r})$.
The part that contains the band projector $\hat{P}_{\bf k+q}$
vanishes identically, which can be seen in the following way.
Any physical scalar field must
be real, which implies
\begin{equation}
n^{(1)}_{\bf -q}({\bf r}) = n^{(1)*}_{\bf q}({\bf r}).
\end{equation}
Thus, we can write the contribution of $\hat{P}_{\bf k+q}$ to $\Delta n^{\beta}_{\bf q}({\bf r})$ as
\begin{equation}
\label{trick}
\begin{split}
\sum_{mj} & i \langle u^{(0)}_{m\bf k}| {\bf r} \rangle \langle {\bf r} | u^{(0)}_{j\bf k+q} \rangle
            \langle u^{(0)}_{j\bf k+q} | \left( \hat{p}_{\bf k\beta} + \frac{q_\beta}{2} \right) | u^{(0)}_{m {\bf k}} \rangle   \\
 - & \sum_{mj} i \langle u^{(0)}_{j\bf k-q}| {\bf r} \rangle \langle {\bf r} | u^{(0)}_{m\bf k} \rangle
             \langle   u^{(0)}_{m {\bf k}}| \left( \hat{p}_{\bf k\beta} - \frac{q_\beta}{2} \right) |u^{(0)}_{j\bf k-q} \rangle.
\end{split}
\end{equation}
After operating a translation in ${\bf k}$-space on the second line (this is irrelevant, as
the expression needs to be integrated over the whole Brillouin zone), the result manifestly
vanishes.

We are left in Eq.~(\ref{deltau}) with just the contribution of the identity operator, which
can be written as
\begin{eqnarray}
 \Delta n^{\beta}_{\bf q}({\bf r}) &=& -\frac{\partial n^{(0)}({\bf r})}{\partial r_\beta} - i q_\beta n^{(0)}({\bf r}),
 \label{rhodiff}
\end{eqnarray}
The form of Eq.~(\ref{rhodiff}) might appear puzzling at first sight, but in fact it
accurately matches the known relationship between the charge-density responses in the
curvilinear and Cartesian reference frames~\cite{Stengel2013natcom,StengelChapter}.
For example, at ${\bf q}=0$, we already know that the metric perturbation (and,
as a consequence, the corresponding density response) must vanish,
\begin{equation}
\hat{\mathcal{H}}^{(\beta)}_{{\bf k,q}=0} = 0, \qquad n^{(\beta)}_{{\bf q}=0}({\bf r}) = 0,
\end{equation}
since a uniform translation of the crystal has no effect in its own co-moving
reference frame.
Also, by translational symmetry one must have, for the laboratory-frame
perturbation
\begin{equation}
\hat{\mathcal{H}}^{u_{\beta}}_{{\bf k,q}=0} = i  \left[  \hat{H}^{(0)}_{\bf k}, \hat{p}_{\bf k \beta} \right],
\end{equation}
which implies that
\begin{equation}
n^{u^{\beta}}_{{\bf q}=0}({\bf r}) = -\frac{\partial n^{(0)}({\bf r})}{\partial r_\beta}.
\end{equation}
One can easily check that our formulas for $\hat{\mathcal{H}}^{(\beta)}_{{\bf k,q}}$,
together with the results Eq.~(\ref{transform}), Eq.~(\ref{rhorelation}) and Eq.~(\ref{rhodiff}),
are fully consistent with
these requirements.

In other words,
Eq.~(\ref{rhodiff}) corroborates our interpretation of the modified perturbation
as a metric wave, where the atomic displacements are expressed as
a local modification of the metric of space.
From this perspective, $|\Delta u^{\beta}_{m {\bf k,q}} \rangle$ is
essential for ensuring that the first-order density response complies with the
established transformation laws.

 \subsection{Implementation Considerations}\label{implem_considerations}

 The formulas derived in the previous section are formal, but when
 implementing them we need to introduce approximations in order to
 make the calculations tractable. In particular, we make a set of
 choices concerning the discrete sampling of the Brillouin zone (BZ)
 with a finite mesh, and the plane-wave energy cutoff used in the
 wave-function expansion.
 It is therefore important to clarify
 which of the above relations remain exact in principle, once such a
 set of choices has been made, and which should be expected to show
 discrepancies (of course, these will diminish as more highly
 converged choices are made).

 Our main focus will be on Eq.~(\ref{rhodiff}), describing the
 difference between the electron-density response to a phonon
 perturbation in the laboratory frame, already available within the
 existing DFPT implementations, and the new metric response introduced
 in this work.

 First of all note that, in order to obtain Eq.~(\ref{rhodiff}), we
 have used the fact that the expression in Eq.~(\ref{trick}) vanishes;
 this, in turn, relies on the fact that it must be integrated over the
 whole Brillouin zone.  If the BZ is sampled by a discrete number of
 ${\bf k}$ points, then Eq.~(\ref{trick}) is only approximately
 satisfied; in fact, one can see that it holds exactly only if the set
 of ${\bf k}$-points is invariant under a translation by ${\bf q}$,
 i.e., ${\bf q}$ is commensurate with the ${\bf k}$-points.

   \begin{figure}
   \includegraphics[width=\columnwidth]{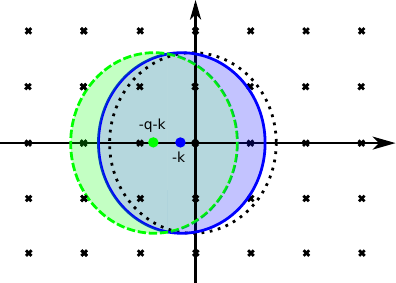}
   \caption{\label{fig:sphere}
             Representation of the Fourier space in 2D.
             The small black crosses are the ${\bf G}$-vectors;
             the black dotted circle identifies the cutoff sphere centered
             on $\Gamma$; the blue continuous circle identifies nonzero
             Fourier coefficients of $u_{\bf k}^{(0)}$;
             and the dashed green circle identifies the Fourier coefficients
             of $\tilde{u}_{\bf k}^{(0)}$.
             }
   \end{figure}

  Next, as we shall see in the following, commensuration between
  ${\bf q}$ and the ${\bf k}$-mesh does not automatically guarantee that
  Eq.~(\ref{rhodiff}) is exact.
To see why, it is useful to write the explicit expression for the charge density
 difference as
\begin{equation}
  \begin{split}
    \label{delrho}
\Delta n^{\beta}_{\bf q} ({\bf r}) &= -\frac{4i}{N_{\bf k}} \sum_{m\bf k} \langle u^{(0)}_{m {\bf k}} |{\bf r} \rangle
  \langle {\bf r}  | \hat{Q}_{\bf k+q}\left( \hat{p}_{\bf k\beta} + \frac{q_\beta}{2} \right) | u^{(0)}_{m {\bf k}} \rangle \\
  & = -\frac{4i}{N_{\bf k}} \sum_{m\bf k} \langle u^{(0)}_{m {\bf k}} |{\bf r} \rangle
  \langle {\bf r}  | \left( \hat{p}_{\bf k\beta} + \frac{q_\beta}{2} \right) | u^{(0)}_{m {\bf k}} \rangle,
\end{split}
\end{equation}
where $N_{\bf k}$ is the number of ${\bf k}$-points (a uniform mesh is
assumed), and the second equality relies on the assumed commensuration
between ${\bf q}$ and the mesh (see above discussion).
Eq.~(\ref{delrho}), however, only satisfies Eq.~(\ref{rhodiff}) in the
limit of a complete plane-wave basis set, i.e., for an infinitely
large plane-wave cutoff, $E_{\rm cut}$. In practice, a finite basis
set is always used, which means that plane waves with a kinetic energy
that is larger than $E_{\rm cut}$ are discarded from the calculation.
Crucially, the kinetic energy of a plane wave is calculated as
$|{\bf G+k}|^2/2$, which implies that different ${\bf k}$ points are
characterized by different cutoff spheres in reciprocal space, and
hence by different basis sets.
For example, the wavefunction $u_{\bf k}^{(0)}({\bf r})$ has non-zero
coefficients only inside a cutoff sphere centered in ${\bf -k}$, while
the sphere of both the phonon and metric response functions,
$u_{\bf k,q}^{(1)}({\bf r})$, is centered in ${\bf -(k+q)}$ (see
Fig.~\ref{fig:sphere}).
Now, note that the function $|\Delta u^\beta_{m\bf k,q}\rangle$ that
we have used to define $\Delta n^{\beta}_{\bf q} ({\bf r})$
``belongs'' to the point ${\bf k+q}$, and hence it will not, in
general, be represented on the same basis set as
$|u^{(0)}_{m\bf k}\rangle$; this is the reason why Eq.~(\ref{rhodiff})
is generally violated when a finite $E_{\rm cut}$ is used.

To illustrate this point more clearly,
we can write the charge-density difference, as it is computed
in practice starting from $n^\beta$ and $n^{u_\beta}$, as
  \begin{equation}
\begin{split}
\Delta n^{\beta}_{\bf q} ({\bf r}) &=
   -\frac{4i}{N_{\bf k}} \sum_{m\bf k} \langle u^{(0)}_{m {\bf k}} |{\bf r} \rangle
  \langle {\bf r}  | 
  \left( \hat{p}_{\bf k\beta} + \frac{q_\beta}{2} \right)  | \tilde{u}^{(0)}_{m {\bf k}} \rangle.
 \label{revised}
\end{split}
\end{equation}
Here $\tilde{u}_{m\bf k}^{(0)}({\bf r})$
is the same as $u_{m\bf k}^{(0)}({\bf r})$ in
Eq.~(\ref{delrho}) except
that it has nonzero Fourier components only in the
intersection between the green and blue circles of
Fig.~\ref{fig:sphere}, while $u_{m\bf k}^{(0)}({\bf r})$
is defined inside the whole blue solid circle.
Since the first-order wavefunctions are obtained
through a self-consistent process, this error will propagate to the
potentials and back to the density; thus, at the end of the
calculation even the ``revised'' relationship Eq.~(\ref{revised}) will
not be exactly fulfilled.
In any case, we can expect that the error will be roughly linear in
$|{\bf q}|$, and should rapidly vanish upon increasing the plane-wave
cutoff; we shall see that both expectations are nicely fulfilled in
our tests.
As we shall show shortly, this discrepancy between the phonon and
metric approach results in a faster numerical convergence of the
latter with respect to plane-wave cutoff and $k$-point sampling.

  \section{Results\label{tests}}

  To test our implementation,
  we have compared the results of the metric perturbation against
  response functions that are already present in publicly available DFPT
  codes: the phonon perturbation~\cite{Abinit_phonon_2} and
  the uniform strain perturbation~\cite{Hamann2005}.
  The quantities that we used to gauge the accuracy
  of the implementation are either based on the
  first-order charge density
  (a fundamental linear-response quantity), or
  on the cell-averaged polarization.
  In particular, we have performed four independent tests:
  \begin{itemize}
  \item In Sec.~\ref{chgphonon}, we compare the electron density response
  of a ``clamped-ion acoustic phonon'' to the electron density
  response to a metric perturbation, following the
  guidelines of Sec.~\ref{subsec:rel_two_frames}.
\item In Sec.~\ref{chgstrain}, we compare the
  electron density response of a uniform strain perturbation to the
  electron density response of a metric perturbation at first order in
  ${\bf q}$.  We have already demonstrated in Eq.~(\ref{Hmetricstain})
  the relation that must hold between the uniform strain Hamiltonian
  and the metric perturbation.
  In the same way, the response density to a metric and the associated
  uniform strain perturbation are related by
  \begin{equation}\label{rhodiffstr}
    - i n^{1,\beta}_\alpha = n_{\alpha\beta}^{\text{strain}},
  \end{equation}
  where $n^{1,\beta}_\alpha$ is the first derivative of
  the microscopic metric perturbation response, $n^{1,\beta}$, respect
  to the $q_{\alpha}$.
  Note that, by time reversal symmetry,
  $n^{1,\beta}_\alpha$ is a pure imaginary function.
 \item In Sec.~\ref{oct}, we compare the octupolar response
  calculated via the phonon to that from  metric perturbation.
  The octupolar tensor components can be extracted via the
  long-wave expansion of the macroscopic (i.e., cell-integrated)
  charge-density response
  \begin{equation}
    Q^{(3,\alpha\beta\gamma)}_\delta = \int_{\rm cell} d^3 r\text{ }
    \left.\frac{\partial^3   n^{(\delta)}_{\bf q}({\bf r})}{\partial q_\alpha \partial q_\beta \partial q_\gamma}
    \right\rvert_{{\bf q}=0}
  \end{equation}
  where $\delta$ indicates the atomic displacement direction.
   Clearly, since  the geometrical term $\Delta n^\beta$
   averages to zero, both the phonon and metric
   calculations must yield the same values of $Q^{(3,\alpha\beta\gamma)}_\delta$.
  The $q$-derivative can be performed by fitting the cell-integrated density
  as a function of  ${\bf q}$ in a vicinity of ${\bf q}=0$, as
  described in Ref.~(\onlinecite{Stengel2013natcom}).
  Testing this quantity is particularly interesting
  in the context of the present
  work because the longitudinal octupole, $Q_{\rm L}=Q^{(3,\alpha\alpha\alpha)}_\alpha$,
  is directly related to the
  longitudinal FxE coefficient by $\mu_{\rm L} = Q_{\rm L}/(6\Omega)$.

\item In Sec.~\ref{flexo_results}, we compare FxE coefficients
  calculated via the phonon method [Eqs.~(\ref{muI}),~(\ref{PqTR0}), and~(\ref{pchi1})] 
  and the metric-wave method [Eqs.~(\ref{PqTR}) and~(\ref{muIprime})].

  \end{itemize}

  Note that, whenever a 3D scalar field is involved (first
  and second tests),
  we shall use the ``distance''
  \begin{equation}\label{errfunc}
    d(f,g) = \frac{1}{\Omega} \int_{\rm cell} d^3 r \, | f({\bf r}) - g({\bf r}) |,
  \end{equation}
  to gauge their overall difference,
  where functions $f$ and $g$ identify the left- and right-hand
  sides of the given relation that is to be verified.

 \subsection{Computational setup}

 We have used two types of systems for our benchmark tests in the
 following sections: isolated noble gas atoms in large boxes, and
 cubic bulk solids.  Regarding the isolated atoms, we have tested
 three different noble gases, He, Ne and Kr.  As for the cubic solids,
 we have used crystalline Si in the diamond structure, and the cubic
 perovskite phase of SrTiO$_3$.

Our calculations are performed in the framework of density-functional
theory, 
using the local-density approximation (we have employed the Perdew-Wang
92 parametrization~\cite{Perdew92} of the exchange and correlation).
The core-valence interactions are described by
Troullier-Martins~\cite{Troullier91} norm-conserving pseudopotentials,
which we have generated via the fhi98PP~\cite{fhi98PP} code with the
following electronic configurations: He=$1s^2$; Ne=$2s^22p^6$;
Kr=$4s^24p^6$; Si=$3s^23p^2$; Sr=$4s^24p^5s^2$; Ti=$3s^23p^63d^24s^2$;
O=$2s^22p^4$.  Note that the He pseudopotetial only contains a local
part.

The noble gas atoms have been simulated in cubic boxes large enough to
avoid interaction between the replicas.  For the tests in
Secs.~\ref{chgphonon}, \ref{chgstrain}, and \ref{oct}, the size of
such box are $5$ Bohr for He, $7$ Bohr for Ne, and $14$ Bohr for Kr,
with (unless specified), Monkhorst-Pack (MP)
$k$-meshes~\cite{Monkhorst76} of 8$\times$8$\times$8 for He and Ne,
and 4$\times$4$\times$4 for Kr.  For the calculation of flexoelectric
constants in Sec.~\ref{flexo_results}, 14 Bohr boxes were used for all
atoms, with a 4$\times$4$\times$4 $k$-mesh and a plane wave cutoff of
120 Ha.

The relaxed cubic lattice parameters obtained for Si and SrTiO$_3$ are
$10.102$ and $7.267$ Bohr respectively. Calculations are performed
under short-circuit electrostatic boundary conditions (see
Refs.~\onlinecite{Stengel2013} and~\onlinecite{Dreyer2018} for
details). For Si and SrTiO$_3$, MP $k$-meshes from
4$\times$4$\times$4 to 16$\times$16$\times$16 and plane wave cutoffs
from 20 to 100 Ha were tested to explore the convergence properties of
the metric and phonon implementations. For the calculations of
flexoelectric coefficients, a 12$\times$12$\times$12 $k$-mesh and 80
Ha plane wave cutoff were used.

 \subsection{Charge density response: Phonon vs. metric \label{chgphonon}}

   \begin{figure}
   \includegraphics[width=\columnwidth]{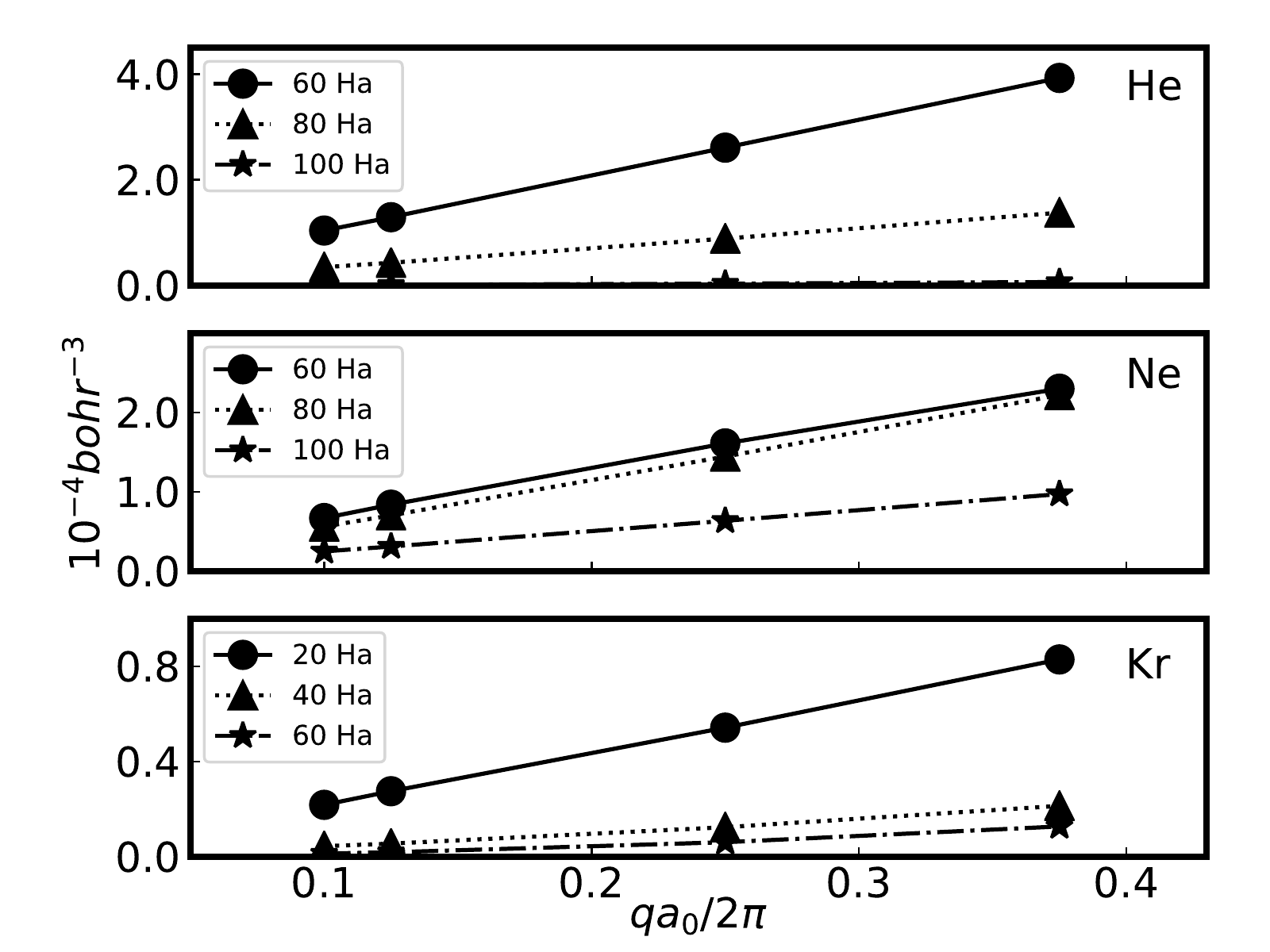}
   \caption{\label{fig3B}
             Plot of
             $d(n^{u^{\beta}}_{\bf q}({\bf r}),
                n^{(\beta)}_{\bf q}({\bf r}) + \Delta n^{\beta}_{\bf q}({\bf r}))$ [cf., Eq.~(\ref{errfunc})]
             as a function of wave vector $\bf q$ (reduced coordinates),
             for different cutoffs.
             All the results refer to longitudinal perturbations.
             From top to bottom: He, Ne and Kr.
             }
   \end{figure}
  First, we check
  the validity of Eq.~(\ref{rhorelation}), which connects the metric
  and phonon charge density response functions via a
  geometric term.
  To make this test quantitative,
  we have taken advantage of the distance function
  defined in Eq.~(\ref{errfunc}),
  with $f({\bf r})=n^{u^{\beta}}_{\bf q}({\bf r})$
  and $g({\bf r}) = n^{(\beta)}_{\bf q}({\bf r}) + \Delta n^{\beta}_{\bf q}({\bf r})$.
  [We construct $\Delta n^{\beta}_{\bf q}({\bf r})$ in terms of the ground-state
  density, following Eq.~(\ref{rhodiff}).]
 Tests are conducted on  He, Ne and Kr
 atoms.
 Due to periodic boundary condition these systems can be regarded as
 crystals of isolated atoms, intended as a computational analog to the
 toy model of Ref.~\onlinecite{Stengel2013natcom}, and discussed
   further in Sec.~\ref{bench}.
  The perturbations considered here are longitudinal, and
  they propagate along
  one of the three equivalent Cartesian axis.
  In Fig.~\ref{fig3B} we report the values of $d(f,g)$
  as a function of the wavevector amplitude, $|\bf q|$, for different
  energy cutoffs.

  The first interesting observation is the almost perfect
  linear trend shown by the function $d(f,g)$.
  As we anticipated in Section \ref{implem_considerations},
  this is a direct consequence of using a finite plane-wave basis set:
  the larger the wavevector, the larger the shift of the cutoff sphere,
  and hence one expects a discrepancy that is roughly proportional to $|\bf q|$.

  Next, one can clearly
  appreciate,  by comparing the slopes of the curves shown in Fig.~\ref{fig3B}, that the
  discrepancy between the phonon
  and metric results decreases as we increase the
  plane-wave cutoff.
  This happens because the discrepancy depends on the
  magnitude of the plane-wave coefficient at the boundary
  of the cutoff sphere (i.e., those falling outside the
  intersection of the two circles in Fig.~\ref{fig:sphere}); this is expected
  to decrease quickly with the cutoff, consistent with our results.
  Also, we see that the discrepancy between the metric and phonon
  charge-density responses is an order of magnitude less for Kr than
  He and Ne; This is a direct consequence of the much softer
  pseudopotential associated to Kr as compared to Ne and He.

  As a final comment we look at the calculated values
  corresponding to wavevectors $\bf q$ that are not
  necessarily commensurate with the $\bf k$-mesh.
  (For example we have used an 8$\times$8$\times$8
  MP $\bf k$-mesh for He and Ne systems, so the point $q=0.1$
  doesn't match the $\bf k$-mesh.)
  The perfect linear trend of the
  distance function for all $\bf q$-values,
  irrespective of the exact or inexact cancellation
  in Eq.~(\ref{trick}) (see discussion in Sec.~\ref{implem_considerations})
  is a clear proof that the $\bf k$-mesh is dense enough,
  so that the finiteness of the plane-wave cutoff is the
  main source of error in this test.

  We stress that the ``discrepancies'' that we discussed above are
  perfectly in line with the expected trends, and thus confirm
  the correctness of the implementation.

  \subsection{Charge density response: Metric vs. uniform strain\label{chgstrain}}

   \begin{figure}
   \includegraphics[width=3.3in]{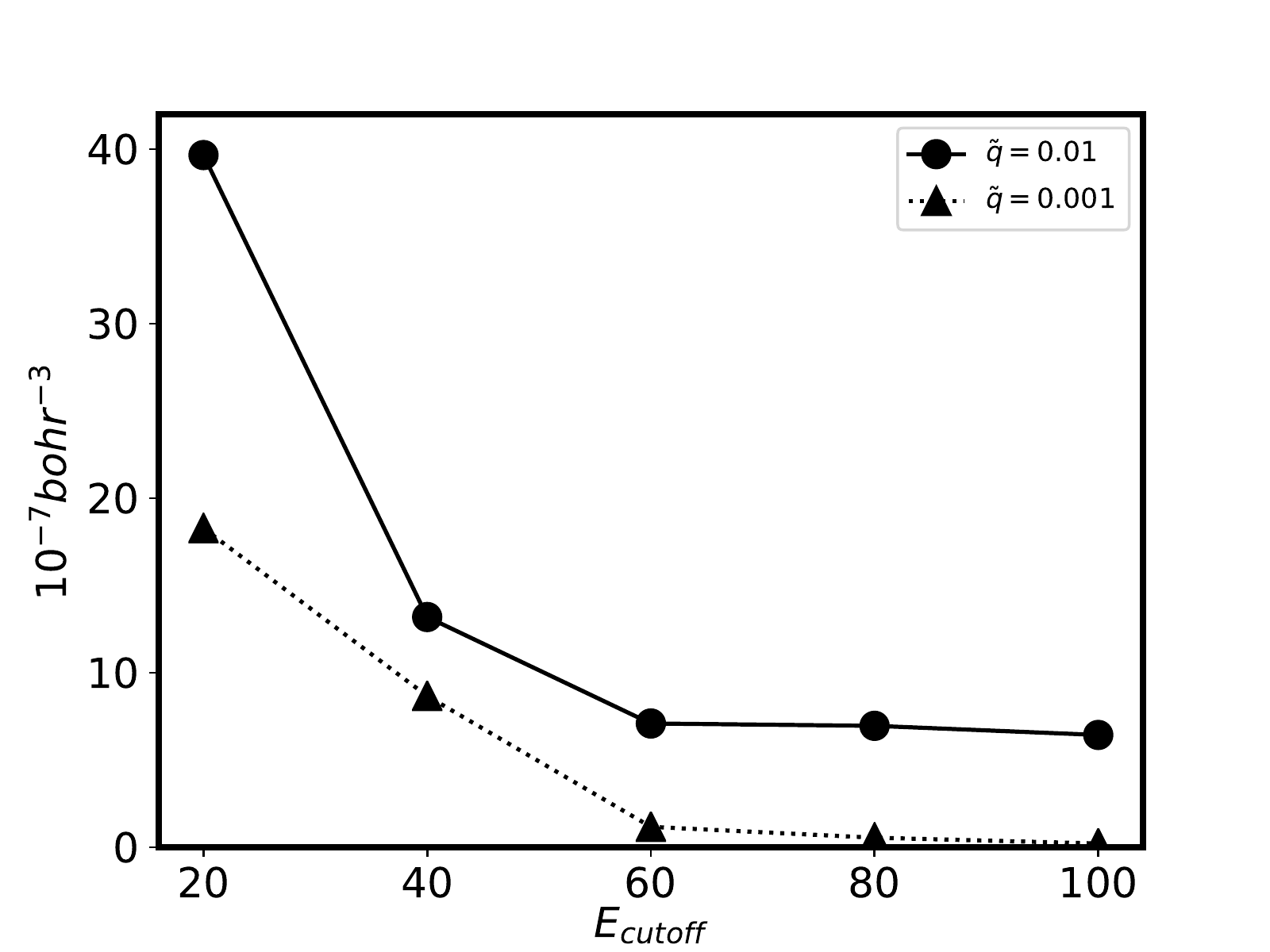}
   \caption{\label{fig:met_vs_str} Cell-integrated difference
     $d(- i n^{1,\beta}_\alpha({\bf r}),n_{\alpha\beta}^{strain}({\bf
       r}))$ between the first-order term of the long-wave expansion
     applied to the metric response density and the uniform strain
     response, calculated as in Ref.~(\onlinecite{Hamann2005}), for a
     He atom in a box undergoing a longitudinal mechanical
     perturbation.  The two curves refer to two different values of the
     finite difference increment used for obtaining $n^{1,\beta}_\alpha({\bf r})$.  }
   \end{figure}

  A second test of the metric implementation is
  based on its relationship with the response to a uniform strain.
  Indeed, the first derivative respect to the wavevector $\bf q$
  of the metric perturbation should coincide with
  the strain perturbation of Hamann \emph{et al.}~\cite{Hamann2005},
  which is already
  implemented in the official release of the
  ABINIT code [see Eq.~(\ref{Hmetricstain})].
  To prove this point, here we use the distance function
  of Eq.~(\ref{errfunc}) to compare the charge-density
  response functions $f({\bf r})=- i n^{1,\beta}_\alpha({\bf r}) $
  and $g({\bf r})= n_{\alpha\beta}^{\text{strain}}({\bf r})$,
  which should coincide according to Eq.~(\ref{rhodiffstr}).
  The derivative respect to $\bf q$ of the metric response
  is performed by finite differences, and using
  two different spacing values: $\Delta q=(2\pi/a_{0})\{0.01;0.001\}$,
  where $a_{0}$ is the
  lattice parameter of the primitive cubic cell.
  In Fig.~\ref{fig:met_vs_str} we show $d(f,g)$
  for the crystal of noninteracting He atoms as function of the energy
  cutoff, $E_{\rm cut}$.
  As expected, the discrepancy rapidly goes to zero
  at larger values of $E_{\rm cut}$, again proving the correctness of the
  implementation.
  We also note that, by reducing the spacing value for the numerical calculation of
  the $\bf q$-derivative,
  the consistency
  between the metric and strain results increases by one order of magnitude.

\subsection{Octupoles\label{oct}}

  \begin{table}
  \caption{\label{TAB1}
           $Q_L/6$ for He and Ne atoms, in $e$ Bohr$^2$
           (Short-circuit electrostatic boundary conditions). }
  \begin{ruledtabular}
  \begin{tabular}{c|cc|cc|cc}
  CutOff & \multicolumn{2}{c|}{He, 5 Bohr}
               & \multicolumn{2}{c|}{Ne, 7 Bohr} & \multicolumn{2}{c}{Kr, 14 Bohr}\\
     (Ha)     & metr & phon & metr & phon & metr & phon\\
  \hline
  40 &  0.4392 & 0.4660 & 1.7338 & 1.7928 & 5.8433 & 5.8382 \\
  60 &  0.4392 & 0.4322 & 1.8129 & 1.8208 & 5.8635 & 5.8618 \\
  80 &  0.4396 & 0.4418 & 1.8135 & 1.8111 & 5.8635 & 5.8635 \\
  100&  0.4398 & 0.4398 & 1.8135 & 1.8135 & 5.8635 & 5.8635 \\
  \end{tabular}
  \end{ruledtabular}
  \end{table}

 \begin{table}
  \caption{\label{TAB2}
           $Q_L/6$ along the [100] direction for Si and along the [110]
           direction for SrTiO$_3$.
           Values are in unit of $e$ Bohr$^2$
           (SC electrostatic boundary conditions).
           The two different columns for Si refer to two different
           $\bf k$-mesh used: 12$\times$12$\times$12 and 16$\times$16$\times$16, respectively. }
 \begin{ruledtabular}
  \begin{tabular}{c|cc|cc|cc}
  CutOff & \multicolumn{2}{c|}{Si(12)} & \multicolumn{2}{c|}{Si(16)} & \multicolumn{2}{c}{STO$_3$} \\
              (Ha)    & metr & phon & metr & phon & metr & phon \\
  \hline
  20 & 478.379 & 478.456 & 478.391 & 478.409 & &  \\
  40 & 478.597 & 478.644 & 478.597 & 478.605 & &  \\
  60 & 478.601 & 478.653 & 478.601 & 478.612 & 111.793 & 111.658 \\
  80 & 478.601 & 478.653 &         &         & 111.662 & 111.666 \\
  100& 478.601 & 478.653 &         &         & 111.684 & 111.673 \\
  \end{tabular}
  \end{ruledtabular}
  \end{table}

  We now compare the longitudinal octupoles
  calculated either using the metric or the
  standard acoustic phonon perturbation.
  Following the procedure described in Ref.~\onlinecite{Stengel2014},
  we interpolate the total density response
  of both the phonon and metric perturbations with a cubic
  polynomial as a function of $\bf q$.
  For this test we have employed the three noble gas atoms (He, Ne, and
  Kr), Si in the diamond structure, and cubic SrTiO$_3$.
  The directions chosen to calculate the longitudinal octupole
  are [100] and [110] for Si and SrTiO$_3$, respectively,
  while the wavevector amplitudes that we use for the
  cubic fit are
  $q_i = \{0.01;0.02;0.03\}$ (in reduced units of $2\pi/a_0$).
  Note that, in the case of the phonon response,
  the electronic charge also has a non-zero linear term
  as a function of $\bf q$, 
  whose slope gives the electronic contribution to the Born effective
  charge of the displaced sublattice.
  Such a linear term is not present in the metric response,
  as the atoms are not moving in the curvilinear frame.

  In Table~\ref{TAB1} we report the values of $Q_{\rm L}/6$ for the
  He, Ne and Kr atoms, while the corresponding
  values for bulk Si and SrTiO$_3$ are shown in Table~\ref{TAB2}.
  As expected, the agreement between the metric and the phonon
    results increases with increasing the plane wave cutoff; such an
  agreement becomes essentially perfect in the case of the isolated
  noble-gas atoms at an energy cutoff of $100$ Ha.
  The metric results converge much
  faster as a function of $E_{\rm cut}$ than the phonon results.
  Moreover, the test between the silicon octupole calculated with a
  12$\times$12$\times$12 and a 16$\times$16$\times$16 MP
  ${\bf k}$-mesh shows that the metric calculation also converges much
  faster with respect to the number of ${\bf k}$-points.

    The relatively worse convergence behavior in
    the phonon case can be tracked down to the quantity
  $\Delta n^\beta_{\bf q}({\bf r})$.
  Indeed, if Eq.~(\ref{rhodiff}) were exactly satisfied, the cell integral of $\Delta n^\beta_{\bf q}({\bf r})$
  would vanish identically, and would not contribute to the calculated octupolar moment.
  However, as we have seen in Sec.~\ref{implem_considerations}, in practical calculations
  Eq.~(\ref{rhodiff}) is violated, and more so at lower energy cutoffs or coarser ${\bf k}$-point
  samplings.
  This can introduce an additional, spurious $\mathcal{O}(q^3)$ contribution to
  the macroscopic density response, and since $\Delta n^\beta_{\bf q}({\bf r})$ is
  rather large, this can have a negative impact on the overall convergence.
  Thus, our numerical tests reveal a further
  (and formerly unexpected) advantage of the metric perturbation
  presented here, i.e., a significant economy in terms of
  computational resources compared with the standard phonon treatment.
  This can be important when dealing with larger systems;
  we shall come back to this point later on.

  \subsection{Flexoelectric coefficients\label{flexo_results}}

  We will now perform calculations of FxE coefficients for our test
  cases, comparing the phonon implementation of
  Ref.~\onlinecite{Dreyer2018} [Eqs.~(\ref{muI}), (\ref{PqTR0}), and
  (\ref{pchi1})] and the metric-wave method implemented in this work
  [Eqs.~(\ref{PqTR}) and (\ref{muIprime})].  
  We will report the flexoelectric tensor components in 
  type-II form (type I and II are
  linearly related to one another),
  following the convention that was established in earlier works.
  Moreover, as we are dealing
  with cubic crystals, the flexoelectric tensor has only three
  independent components,
  which are indicated as the longitudinal
  ($\mu_{\text{L}}\equiv\mu_{11,11}$), transverse
  ($\mu_{\text{T}}\equiv\mu_{11,22}$), and shear
  ($\mu_{\text{S}}\equiv\mu_{12,12}$) flexoelectric coefficients
  henceforth (also remind that in cubic crystals 
  the diamagnetic susceptibility is isotropic,
  $\chi_{\gamma\lambda}^{\text{mag}}=\delta_{\gamma\lambda}\chi^{\text{mag}}$).
  In all cases, the second derivatives with respect
  to \textbf{q} necessary for Eqs.~(\ref{muI}) and (\ref{muIprime})
  will be taken numerically with $\Delta q=(2\pi/a_0)0.005$.

\subsubsection{Isolated spherical atoms\label{bench}}

In order to test the metric implementation for calculating FxE
coefficients, we consider the toy model of a material made of isolated
(i.e., noninteracting), spherical charge densities that was explored
in Refs.~\onlinecite{quantum},~\onlinecite{Stengel2013natcom},
and~\onlinecite{StengelChapter}.  In earlier works, this was denoted
as the isolated rigid charge (IRC) model, and we shall follow such
naming convention here.
Of course, such a material is fictitious, since it would have no
interatomic forces to hold it together. However, it serves as an
interesting test case, since its FxE properties can be determined
analytically and compared to our numerical calculations.  As before,
\cite{Dreyer2018} we can approximate this model by performing DFT
calculations on noble gas atoms in large enough simulation cells
(filled with vacuum) that they do not interact with their periodic
images. As pointed out in
Ref.~\onlinecite{Dreyer2018}, however, in practical calculations atoms
are not ``rigid'' but slightly polarizable, and the model needs to be
revised to account for this fact.

Based on the revised IRC model, for the metric implementation
(under SC boundary conditions) we expect that the flexoelectric
coefficients will be \cite{Dreyer2018,quantum}
\begin{equation}
\label{muNG1}
 \mu^\prime_{\text{NG,L}}=\mu^\prime_{\text{NG,T}}=\epsilon\frac{Q_{\text{NG}}}{2\Omega},
 \end{equation}
and
 \begin{equation}
\label{muNG2}
\mu^\prime_{\text{NG,S}}=0,
\end{equation}
where the subscript ``NG'' indicates a DFT calculation on a noble gas
atom, $\epsilon=\delta_{ij}\epsilon_{ij}$ is the isotropic clamped-ion
dielectric constant, and $Q_{\text{NG}}$ is the quadrupole moment of
the ground-state charge density of the noble-gas atom. The presence of
$\epsilon$ accounts for the fact that the noble gas atoms (in contrast
to the IRC model charge densities) are slightly
polarizable,\cite{Dreyer2018} as we mentioned above.

 Table \ref{IRCtab} gives the clamped-ion FxE coefficients calculated
for noble gas atoms using the metric and phonon
implementations. By comparing the $\mu^\prime_{\text{L}}$,
$\mu^\prime_{\text{T}}$, and $\epsilon Q_{\text{NG}}/2\Omega$ columns,
we see that Eq.~(\ref{muNG1}) is satisfied to within our numerical
accuracy for both the metric and phonon methods. In addition,
$\mu^\prime_{\text{S}}$ vanishes (within our numerical accuracy). The
main source of error is the numerical differentiation of the induced
polarization with \textbf{q} in order to obtain Eqs.~(\ref{muI}) and
(\ref{muIprime}). The results of Table \ref{IRCtab} indicate that the
metric implementation is an accurate method for calculating
flexoelectric coefficients, with increased efficiency as discussed
above.

\begin{table}
  \caption{Clamped-ion flexoelectric coefficients calculated using the metric and phonon 
  implementations, as well as the quadrupole moments of the ground-state charge density. In the 
  case of the phonon perturbation the dynamic contribution, proportional to $\chi^{\text{mag}}$,
  has been removed. 
All quantities are in units of pC/m. }
\begin{ruledtabular}
\label{IRCtab}
\begin{tabular}{c|cc|cc|c|r@{.}lr@{.}l}
&\multicolumn{2}{c|}{$\mu^\prime_{\text{L}}$}  &\multicolumn{2}{c|}{$\mu^\prime_{\text{T}}$}&&\multicolumn{3}{c}{$\mu^\prime_{\text{S}}\text{  }(10^{-3})$} \\
& metr & phon & metr & phon &$\epsilon Q_{\text{NG}}/2\Omega$& \multicolumn{2}{c}{metr} & \multicolumn{2}{c}{phon}  \\
\hline
  He &$-0.479$ &$-0.479$& $-0.479$&$-0.479$&$-0.479$&  $0$&$0$ &$-0$&$3$ \\
 Ne &$-1.843$&$-1.844$&$-1.841$&$-1.842$&$-1.842$&    $-0$&$7$ &$-0$&$6$ \\
 Kr &$-6.477$ &$-6.470$&$-6.476$  &$-6.476$&$-6.479$& $-0$&$3$ &$-0$&$5$\\
 \end{tabular}
\end{ruledtabular}
\end{table}

\subsubsection{Cubic materials\label{cubic}}

We will now apply the metric implementation to calculate the bulk,
clamped-ion FxE coefficients for two prototypical materials: SrTiO$_3$
(in the high-temperature cubic phase) and Si. As with the isolated
atoms, we calculate the primed FxE coefficients from the phonon method
by calculating $\chi^{\text{mag}}$ and using Eq.~(\ref{pchi1}); the
metric implementation gives us the prime coefficients directly. We can
see from Table \ref{Cubtab} that the agreement between the metric and
phonon implementations is excellent.

As observed in previous calculations of the clamped-ion FxE
coefficients \cite{Stengel2014,Dreyer2018}, we see that
$\mu^\prime_{\text{L}}\simeq \mu^\prime_{\text{T}}\gg
\mu^\prime_{\text{S}}$, which is similar to the behavior of the
isolated atoms. However, a distinct difference is the importance of
the dynamic contribution. We saw in Sec.~\ref{bench} that for the
isolated atoms, $\chi^{\text{mag}}$ was the same order as the
longitudinal and transverse coefficients, whereas in the case of the
cubic materials, the $\chi^{\text{mag}}$ is two orders of magnitude
smaller.

In spite of the small magnitude of $\chi^{\text{mag}}$, our results
are sufficiently converged to see clearly that the
rotation-gradient correction is required for accurate agreement
between the metric and phonon implementations. If we neglect this
correction, i.e., calculating $\mu$ with the phonon approach instead
of $\mu^\prime$ [see Eq.~(\ref{pchi1})], we obtain
$\mu_{\text{T}}=-0.810$ and $\mu_{\text{S}}=-0.091$ for SrTiO$_3$, and
$\mu_{\text{T}}=-1.070$ and $\mu_{\text{S}}=-0.180$ for Si, which have
clear discrepancies with the metric results in Table \ref{Cubtab}.

\begin{table}
  \caption{Flexoelectric constants for SrTiO$_3$ and Si calculated using the phonon and metric
   implementations (units are nC/m); their orbital magnetic susceptibilities, $\chi^{\text{mag}}$,
   are respectively
  $-8.3\times 10^{-3}$ and $10.2\times 10^{-3}$nC/m. }
\begin{ruledtabular}
\label{Cubtab}
\begin{tabular}{c|cc|cc|cc}
& \multicolumn{2}{c|}{$\mu^\prime_{\text{L}}$}&\multicolumn{2}{c|}{$\mu^\prime_{\text{T}}$}&\multicolumn{2}{c}{$\mu^\prime_{\text{S}}$}  \\
&metr &phon&metr &phon&metr &phon\\
\hline
 SrTiO$_3$ &$-0.885$&$-0.884$&$-0.826$&$-0.826$&$-0.082$&$-0.083$\\
 Si  &$-1.411$&$-1.410$&$-1.049$&$-1.050$&$-0.189$&$-0.190$\\
\end{tabular}
\end{ruledtabular}
\end{table}

\begin{figure}
   \includegraphics[width=\columnwidth]{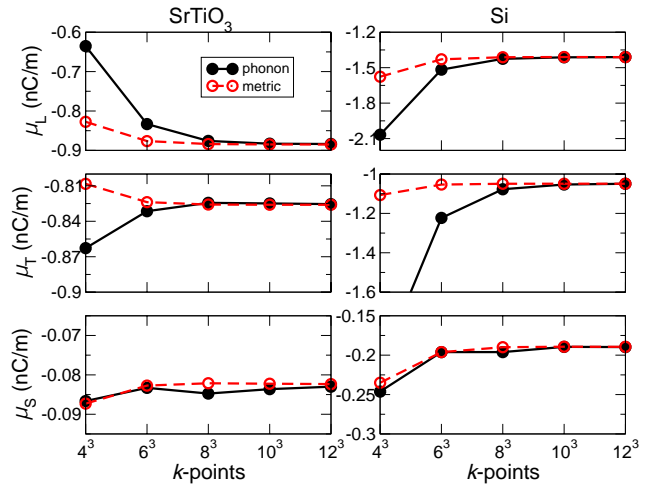}
   \caption{\label{fig:fxekcon}
       Convergence of the FxE coefficients of SrTiO$_3$ and Si with $k$-point mesh for the phonon and metric implementations.
             }
   \end{figure}

In Figs.~\ref{fig:fxekcon} we show the convergence of the FxE
coefficients of SrTiO$_3$ and Si as a function of $k$-point mesh. We
find that the metric implementation shows significantly more rapid
convergence than the phonon implementation (they have similar
  convergence behavior with respect to plane-wave cutoff).
  The slower convergence of the phonon approach may have several
  possible origins. First, there may be additional numerical errors
  associated with the separate calculation of $\chi^{\text{mag}}$ (see
  Eq.~\ref{pchi1}) that is needed for the phonon implementation but
  not for the metric.
Also, as mentioned in Sec.~\ref{metres}, the expansion of the nonlocal
contribution to the current density operator in the case of the metric
implementation can be truncated to a lower order in ${\bf q}$ than in
the phonon case.
Finally, the two implementations differ with respect to the behavior
of the local potential at $\textbf{q}=0$, as will be discussed in
Sec.~\ref{disc}.

   These calculations clearly demonstrate the superiority of the
   metric implementation for determining the clamped-ion FxE
   coefficients. For the case of SrTiO$_3$, for example, a calculation
   of the induced transverse polarization for a perturbation of a
   given \textbf{q} [i.e, Eqs.~(\ref{PqTR0}) and (\ref{PqTR})] using
   the metric implementation took less than 17\% of the cpu time of
   the phonon implementation, mostly because separate calculations for
   the different sublattices was not required. Additional savings in
   the calculation of the FxE coefficient also come from the fact that
   a calculation of $\chi^{\text{mag}}$ is not required for the metric
   implementation.

\subsection{Discussion \label{disc}}

 Here we discuss a series of technical points
 related to the metric perturbation, together with
 possible future generalizations to other physical properties.

 The first observation concerns
 the behavior of the ``external potential'' in the limit
 of ${\bf q} \rightarrow 0$.
 It is well known that the phonon perturbation
 diverges therein, and such divergence is carried by the
 ${\bf G}=0$ Fourier component of the local
 potential (sum over sublattices of Eq.~(\ref{vloc})),
 \begin{equation}
   \sum_\kappa V_{\bf q}^{\text{loc},\tau_{\kappa\beta}} ({\bf G}= 0) = -i q_\beta \frac{1}{\Omega} \sum_\kappa v_\kappa^{\text{loc}}({\bf q}),
 \end{equation}
 where the contribution of each individual sublattice goes like
 \begin{equation}
 v_\kappa^{\text{loc}}({\bf q}) \sim 4 \pi \frac{Z_\kappa}{q^2}.
 \end{equation}
(Recall that $Z_\kappa$ is the total pseudopotential charge.)
The local pseudopotential contribution to the metric perturbation, Eq.~(\ref{Vloc}),
is characterized by an analogous divergence, but the latter is exactly
canceled by an equal and opposite divergence in the geometric
Hartree term, Eq.~(\ref{VH0}),
 \begin{equation}
 \begin{split}
   V_{{\bf q}}^{\text{H0},(\beta)}({\bf G}= 0) &= 4\pi i \frac{q_{\beta}}{q^2} n^{(0)}({\bf G}=0)
 \end{split}
 \end{equation}
 [recall that 
 $n^{(0)}({\bf G}=0)=(1/\Omega)\sum_\kappa Z_\kappa$, as the
 cell must be overall charge-neutral].
 The fact that $\hat{H}^{(\beta)}_{\bf k,q}$ remains finite (in fact, it vanishes)
 in the limit ${\bf q} \rightarrow 0$ might also help explain the superior numerical
 behavior of the metric perturbation in the convergence tests.

 The second point we want to stress concerns the electrostatic boundary conditions
 used in the present work.
 Throughout this work we have implicitly adopted
 short circuit electrostatic boundary conditions (EBC),
 because we were mainly interested in long-wave expansions
 of the polarization (or charge-density) response;
 to do this, it is essential to deal with an analytic function, and
 the short circuit EBC precisely remove the non-analyticity generated by the
 presence of the macroscopic electric field~\cite{Stengel2013}.
 This, however, differs from the physical electrostatic
 conditions (``mixed'' EBC) that characterize the phonon response
 at nonzero $\bf q$~\cite{Hong2013}.
 (The difference between the two cases only concern the longitudinal
 deformations, since with the mixed EBC the mechanical deformation generates a
 macroscopic electric field.)
 Thus, if the metric perturbation is to be employed for the realistic
 simulation of a finite-${\bf q}$ acoustic phonon, such longitudinal
 fields must be incorporated in the calculation~\cite{Stengel2014}.
 For  a metric wave the short circuit EBC are
 obtained by simply removing the ${\bf G}=0$
 component from the self-consistent part of the first-order
 Hartree potential response, $\hat{V}_{\bf q}^{(\beta)}$; by
 plugging this contribution back into the first-order Hamiltonian, we readily recover
 the correct electrostatics.
 Thus, switching from short-circuit to standard electrical
 boundary conditions is even simpler in the metric case than in the standard
 phonon case.

 We conclude this subsection by briefly sketching other possible
 applications of the metric wave perturbation.
 In the present work we have focused
 our attention on two response functions to the metric wave: the
 charge density (Eq.~\ref{rhodiff}) and the electronic polarization
 (Eq.~\ref{PqTR}).
 However, the knowledge of the wave functions $\left\lvert u^{(\beta)}_{n \bf k,q} \right\rangle$
 can be used to calculate other useful physical quantities.
 An obvious candidate is the force-response tensor, defined by the
 force induced on the individual atomic sublattices by an acoustic phonon propagating along ${\bf q}$.
 This can be used~\cite{Stengel2013} to calculate the lattice-mediated contributions to the
 flexoelectric tensor via an appropriate long-wave expansion in ${\bf q}$.
 To see how, we can use a similar strategy as in the electronic polarization case,
 by exploiting the connection between the phonon and metric response functions.
 For example, we can write the ``variational'' contribution to the force experienced
 by the sublattice $\kappa$ in direction $\beta$ induced by an acoustic phonon
 as~\cite{Gonze1997}
 \begin{equation}
 f_{\kappa\beta,\alpha}^{(\rm var,\bf q)}= \frac{4}{N_\kappa}\sum_{n\bf k} \langle  u^{u_\alpha}_{n \bf k,q} \vert H^{\tau_{\kappa\beta}}_{\bf k,q} \vert u_{n \bf k}^{(0)}\rangle,
 \end{equation}
 where
 $|u^{u_\alpha}_{n \bf k,q}\rangle = \sum_{\kappa'} |u^{\tau_{\kappa' \alpha}}_{n \bf k,q}\rangle$
is the response to an acoustic phonon in the laboratory frame, defined as usual as
the sublattice sum of individual atomic displacement.
 Then $|u^{u_\alpha}_{n \bf k,q}\rangle$ can be replaced, by using Eq.~(\ref{umet}),
 with the metric response function, $|u^{(\alpha)}_{n \bf k,q}\rangle$,
 plus an additional piece that only depends on the ground-state wavefunctions, and
 can therefore be reabsorbed into the ``nonvariational'' part.
 Similar considerations could be used, in principle,
 to get the acoustic
 activity tensor~\cite{Portigal1968}
 or the strain-gradient elastic tensor~\cite{Stengel2016}, which
 correspond to third and fourth order in ${\bf q}$ of
 the metric-metric response.

\section{Conclusions\label{con}}

 In this work we have implementated and
 tested a new metric wave perturbation, defined as an acoustic phonon
 described in the frame that is co-moving with the atoms,
  in the context of DFPT.
 It is aimed at calculating the physical response of a crystalline
 material to a generic mechanical deformation, and formally bridges
 the gap between the already available ``phonon''~\cite{Gonze1997}
 and ``uniform strain''~\cite{Hamann2005} perturbations.
 By focusing on the calculation of the flexoelectric tensor
 components, we have demonstrated, via extensive numerical validation,
 its clear advantages in terms of efficiency and
 ease of use with respect to earlier approaches.
 We also study its convergence properties with respect to
 various computational parameters, and find them to be
 very favorable. We rationalize this finding by comparing
 (both analytically and numerically) the charge-density response
 to the metric and standard phonon perturbations.
 We anticipate that, going forward, the methodology presented
 here can become a standard approach for the first-principles
 computation of flexoelectric and related properties of materials.

 \begin{acknowledgments}

 The Flatiron Institute is a division of the Simons Foundation.
 C.~E.~D. and D.~V. was supported by ONR Grant N00014-16-1-2951.
 A.~S. and M.~S. acknowledge the support of Ministerio de Economia,
 Industria y Competitividad (MINECO-Spain) through
 Grants  
 No.  MAT2016-77100-C2-2-P  and  No.  SEV-2015-0496,
 and  of Generalitat de Catalunya (Grant No. 2017 SGR1506).
 This project has received funding from the European
 Research Council (ERC) under the European Union's
 Horizon 2020 research and innovation program (Grant
 Agreement No. 724529). Part of the calculations were performed at
 the Supercomputing Center of Galicia (CESGA).
\end{acknowledgments}

\appendix

\section{Derivation of the pseudopotential terms}
\label{app:pseudo}

Here we carry out explicitly the derivation of the
first-order pseudopotential
terms in curvilinear coordinates, Eq.s~(\ref{Vloc}) and~(\ref{Vsep}).
The curvilinear coordinates are defined by Eq.~(\ref{coord}).
Following the results of Ref.~(\onlinecite{quantum}), a generic (non-local)
pseudopotential operator in the curvilinear coordinates is
\begin{equation}
\tilde{V}^{\rm psp,(0)}(\bm{\xi},\bm{\xi}') = \sqrt{h(\bm{\xi})} \, V^{\rm psp,(0)}({\bf r}(\bm{\xi}),{\bf r}(\bm{\xi}')) \, \sqrt{h(\bm{\xi}')},
\end{equation}
where $h(\bm{\xi})$ is the determinant of the deformation gradient
tensor, $h_{\alpha\beta}=\frac{\partial x_\alpha}{\partial \xi_\beta}$, that in the linear approximation is
\begin{equation}
  h(\bm{\xi})=1+i{\bf q}\cdot \bm{\lambda} e^{i \bm{\xi}\cdot {\bf q}}
\end{equation}

\subsection{Local potential}

By using the transformation properties of the Dirac delta
(see local term in Eq.~(\ref{Vlocsep})), one can easily verify that the
factors of $\sqrt{h}$ cancel out in the local part,
\begin{equation}
\tilde{V}^{\rm loc,(0)}(\bm{\xi}) = V^{\rm loc,(0)} ({\bf r}(\bm{\xi})).
\end{equation}
Using Eqs.~(\ref{Vloccart}) and (\ref{coord}), this immediately leads to
\begin{eqnarray}
\tilde{V}^{\rm loc,(0)}(\bm{\xi}) &=&
\sum_{l\kappa} v_{\kappa}^{\rm loc} [\bm{\xi} - {\bf R}_{l\kappa} +
\bm{\lambda} \, (  e^{i  \bm{\xi}\cdot {\bf q} } - e^{i {\bf R}_{l\kappa} \cdot {\bf q} })
] \nonumber \\
 &=&  V^{\rm loc,(0)} (\bm{\xi}) +   
  e^{i  \bm{\xi}\cdot {\bf q} } \, \sum_{l\kappa} [ 1- e^{i ({\bf R}_{l\kappa} - \bm{\xi}) \cdot {\bf q} }] \times  \nonumber \\
   &&                 \bm{\lambda} \cdot \bm{\nabla} v_{\kappa}^{\rm loc} (\bm{\xi} - {\bf R}_{l\kappa}).
\end{eqnarray}
 and therefore, for the cell-periodic part of the first-order contribution,
\begin{equation}
V_{\bf q}^{{\rm loc}, (\beta)}(\bm{\xi}) = \sum_{l\kappa} [ 1- e^{i ({\bf R}_{l\kappa} - \bm{\xi}) \cdot {\bf q} }] \,
                     \frac{\partial }  {\partial \xi_\beta} v_{\kappa}^{loc} (\bm{\xi} - {\bf R}_{l\kappa})  .
\end{equation}
Note the fact that the first-order potential vanishes identically at ${\bf q}=0$, which is
a consequence of adopting the curvilinear reference system.

To evaluate the Fourier transform it is useful to bring the derivative sign out of the
lattice sum in $V_{\bf q}^{{\rm loc}, (\beta)}(\bm{\xi})$, obtaining the following three pieces,
\begin{eqnarray}
V_{\bf q}^{{\rm loc}, (\beta)}(\bm{\xi}) &=&
\frac{\partial}{\partial \xi_\beta } \sum_{l\kappa}  v^{\rm loc}_\kappa ( \bm{\xi}-  {\bf R}_{l\kappa})  \nonumber \\
 && - \frac{\partial}{\partial \xi_\beta } \left\{ \sum_{l\kappa} e^{i {\bf q} \cdot ({\bf R}_{l\kappa}-\bm{\xi})}
  v^{\rm loc}_\kappa ( \bm{\xi}-  {\bf R}_{l\kappa}) \right\}  \nonumber \\
 && - i q_\beta  \sum_{l\kappa}
  e^{i {\bf q} \cdot ({\bf R}_{l\kappa}-\bm{\xi})} v^{\rm loc}_\kappa ( \bm{\xi}-  {\bf R}_{l\kappa})
\end{eqnarray}
By defining (Ref.~\onlinecite{Gonze1997}, Eq.~(A16)) the Fourier transform of the local atomic potential,
\begin{equation}
v^{\rm loc}_\kappa({\bf K}) = \int d^3 r \, e^{-i{\bf K \cdot r}} v^{\rm loc}_\kappa({\bf r}),
\end{equation}
we can readily evaluate the reciprocal-space expression
for the perturbed
local potential, which is precisely
Eq.~(\ref{Vloc}).

\subsection{Separable potential}

To evaluate the separable part, first recall that
\begin{eqnarray}
\sqrt{h} &=&  1 + \frac{i}{2} \bm{\lambda} \cdot {\bf q} \, e^{i \bm{\xi} \cdot {\bf q} }, \\
\zeta_{\mu \kappa}({\bf r}(\bm{\xi}) - {\bf r}({\bf R}_{l\kappa})) &=& \zeta_{\mu \kappa}(\bm{\xi}-{\bf R}_{l\kappa}) \, + \nonumber \\
  &&  e^{i  \bm{\xi}\cdot {\bf q} } \,  [ 1- e^{i ({\bf R}_{l\kappa} - \bm{\xi}) \cdot {\bf q} }] \times  \nonumber \\
           &&          \bm{\lambda} \cdot \bm{\nabla} \zeta_{\mu \kappa} (\bm{\xi} - {\bf R}_{l\kappa}).
\end{eqnarray}
It is also useful to remind some basic properties of the Fourier transformation of separable
operators. Assume that we wish to express, in Fourier space, the following cell-periodic function
\begin{equation}
F({\bf r},{\bf r}') = \sum_l f({\bf r}-{\bf R}_{l\kappa}) g^*({\bf r}'-{\bf R}_{l\kappa}).
\end{equation}
We have, following  Eq.~(A19) of Ref.~(\onlinecite{Gonze1997}),
\begin{eqnarray}
F({\bf G+k},{\bf G'+k}) &=& \frac{1}{\Omega}  e^{ i ({\bf G'-G}) \cdot \bm{\tau}_\kappa } f({\bf G+k}) g^* ({\bf G'+k}).
  \nonumber \\
\end{eqnarray}
Another basic relationship that we need is
\begin{equation}
\int d^3 r \, f^*({\bf r}) e^{i {\bf K}' \cdot {\bf r} } =
  \Big( \int d^3 r \, f({\bf r}) e^{i {\bf K}' \cdot {\bf r} } \Big)^* = f^*({\bf K}').
\end{equation}
We get, after some algebra,
\begin{equation}
V^{{\rm sep},(\beta)}_{\bf k,q}({\bf G},{\bf G}') = \frac{1}{\Omega} \sum_{\kappa \mu} e_{\mu \kappa} e^{i ({\bf G'-G}) \cdot \bm{\tau}_\kappa }
  f_{\kappa \mu}^{(\beta,{\bf q})} ({\bf K},{\bf K}'),
\end{equation}
with (the first two terms come from the volume factors, third and fourth from the linear variation of $\zeta$,
fifth and sixth from $\zeta^*$)
\begin{eqnarray}
f_{\kappa \mu}^{(\beta,{\bf q})} ({\bf K},{\bf K}') &=& \frac{i}{2} q_\beta \, \zeta({\bf K}) \, \zeta^*({\bf K}') \nonumber \\
  &&       + \, \frac{i}{2} q_\beta \, \zeta({\bf K+q}) \, \zeta^*({\bf K'+q}) \nonumber \\
  &&       + \, i K_\beta \, \zeta({\bf K}) \, \zeta^*({\bf K}')   \nonumber \\
     && -i (K_\beta +q_\beta)  \,  \zeta({\bf K + q})  \, \zeta^*({\bf K}')  \nonumber \\
     && -i (K'_\beta + q_\beta) \,  \zeta({\bf K+q})  \,  \zeta^*({\bf K'+q}) \nonumber \\
 && + \, i K'_\beta  \, \zeta({\bf K + q}) \, \zeta^*({\bf K'})
\end{eqnarray}
%
This expression can be further simplified as follows,
\begin{eqnarray}
f_{\kappa \mu}^{(\beta,{\bf q})} ({\bf K},{\bf K}') &=& i( K_\beta + \frac{q_\beta}{2}) \, \zeta({\bf K}) \, \zeta^*({\bf K}') \nonumber \\
  &&       -i  ( K'_\beta + \frac{q_\beta}{2}) \, \zeta({\bf K+q}) \, \zeta^*({\bf K'+q}) \nonumber \\
     && -i (K_\beta - K'_\beta +q_\beta)  \,  \zeta({\bf K + q})  \, \zeta^*({\bf K}'). \nonumber \\
\end{eqnarray}
The final result is precisely Eq.~(\ref{Vsep})

\bibliography{article}

\begin{thebibliography}{29}%
\makeatletter
\providecommand \@ifxundefined [1]{%
 \@ifx{#1\undefined}
}%
\providecommand \@ifnum [1]{%
 \ifnum #1\expandafter \@firstoftwo
 \else \expandafter \@secondoftwo
 \fi
}%
\providecommand \@ifx [1]{%
 \ifx #1\expandafter \@firstoftwo
 \else \expandafter \@secondoftwo
 \fi
}%
\providecommand \natexlab [1]{#1}%
\providecommand \enquote  [1]{``#1''}%
\providecommand \bibnamefont  [1]{#1}%
\providecommand \bibfnamefont [1]{#1}%
\providecommand \citenamefont [1]{#1}%
\providecommand \href@noop [0]{\@secondoftwo}%
\providecommand \href [0]{\begingroup \@sanitize@url \@href}%
\providecommand \@href[1]{\@@startlink{#1}\@@href}%
\providecommand \@@href[1]{\endgroup#1\@@endlink}%
\providecommand \@sanitize@url [0]{\catcode `\\12\catcode `\$12\catcode
  `\&12\catcode `\#12\catcode `\^12\catcode `\_12\catcode `\%12\relax}%
\providecommand \@@startlink[1]{}%
\providecommand \@@endlink[0]{}%
\providecommand \url  [0]{\begingroup\@sanitize@url \@url }%
\providecommand \@url [1]{\endgroup\@href {#1}{\urlprefix }}%
\providecommand \urlprefix  [0]{URL }%
\providecommand \Eprint [0]{\href }%
\providecommand \doibase [0]{http://dx.doi.org/}%
\providecommand \selectlanguage [0]{\@gobble}%
\providecommand \bibinfo  [0]{\@secondoftwo}%
\providecommand \bibfield  [0]{\@secondoftwo}%
\providecommand \translation [1]{[#1]}%
\providecommand \BibitemOpen [0]{}%
\providecommand \bibitemStop [0]{}%
\providecommand \bibitemNoStop [0]{.\EOS\space}%
\providecommand \EOS [0]{\spacefactor3000\relax}%
\providecommand \BibitemShut  [1]{\csname bibitem#1\endcsname}%
\let\auto@bib@innerbib\@empty
\bibitem [{\citenamefont {Zubko}\ \emph {et~al.}(2007)\citenamefont {Zubko},
  \citenamefont {Catalan}, \citenamefont {Buckley}, \citenamefont {Welche},\
  and\ \citenamefont {Scott}}]{Zubko07}%
  \BibitemOpen
  \bibfield  {author} {\bibinfo {author} {\bibfnamefont {P.}~\bibnamefont
  {Zubko}}, \bibinfo {author} {\bibfnamefont {G.}~\bibnamefont {Catalan}},
  \bibinfo {author} {\bibfnamefont {A.}~\bibnamefont {Buckley}}, \bibinfo
  {author} {\bibfnamefont {P.}~\bibnamefont {Welche}}, \ and\ \bibinfo {author}
  {\bibfnamefont {J.}~\bibnamefont {Scott}},\ }\href@noop {} {\bibfield
  {journal} {\bibinfo  {journal} {Phys. Rev. Lett.}\ }\textbf {\bibinfo
  {volume} {99}},\ \bibinfo {pages} {167601} (\bibinfo {year}
  {2007})}\BibitemShut {NoStop}%
\bibitem [{\citenamefont {Majdoub}\ \emph {et~al.}(2008)\citenamefont
  {Majdoub}, \citenamefont {Sharma},\ and\ \citenamefont {Cagin}}]{Majdoub08}%
  \BibitemOpen
  \bibfield  {author} {\bibinfo {author} {\bibfnamefont {M.}~\bibnamefont
  {Majdoub}}, \bibinfo {author} {\bibfnamefont {P.}~\bibnamefont {Sharma}}, \
  and\ \bibinfo {author} {\bibfnamefont {T.}~\bibnamefont {Cagin}},\
  }\href@noop {} {\bibfield  {journal} {\bibinfo  {journal} {Phys. Rev. B}\
  }\textbf {\bibinfo {volume} {77}},\ \bibinfo {pages} {125424} (\bibinfo
  {year} {2008})}\BibitemShut {NoStop}%
\bibitem [{\citenamefont {Narvaez}\ \emph {et~al.}(2016)\citenamefont
  {Narvaez}, \citenamefont {Vasquez-Sancho},\ and\ \citenamefont
  {Catalan}}]{Narvaez16}%
  \BibitemOpen
  \bibfield  {author} {\bibinfo {author} {\bibfnamefont {J.}~\bibnamefont
  {Narvaez}}, \bibinfo {author} {\bibfnamefont {F.}~\bibnamefont
  {Vasquez-Sancho}}, \ and\ \bibinfo {author} {\bibfnamefont {G.}~\bibnamefont
  {Catalan}},\ }\href@noop {} {\bibfield  {journal} {\bibinfo  {journal}
  {Nature}\ }\textbf {\bibinfo {volume} {538}},\ \bibinfo {pages} {219}
  (\bibinfo {year} {2016})}\BibitemShut {NoStop}%
\bibitem [{\citenamefont {Bhaskar}\ \emph {et~al.}(2016)\citenamefont
  {Bhaskar}, \citenamefont {Banerjee}, \citenamefont {Abdollahi}, \citenamefont
  {Wang}, \citenamefont {Schlom}, \citenamefont {Rijnders},\ and\ \citenamefont
  {Catalan}}]{Bhaskar2016}%
  \BibitemOpen
  \bibfield  {author} {\bibinfo {author} {\bibfnamefont {U.~K.}\ \bibnamefont
  {Bhaskar}}, \bibinfo {author} {\bibfnamefont {N.}~\bibnamefont {Banerjee}},
  \bibinfo {author} {\bibfnamefont {A.}~\bibnamefont {Abdollahi}}, \bibinfo
  {author} {\bibfnamefont {Z.}~\bibnamefont {Wang}}, \bibinfo {author}
  {\bibfnamefont {D.~G.}\ \bibnamefont {Schlom}}, \bibinfo {author}
  {\bibfnamefont {G.}~\bibnamefont {Rijnders}}, \ and\ \bibinfo {author}
  {\bibfnamefont {G.}~\bibnamefont {Catalan}},\ }\href@noop {} {\bibfield
  {journal} {\bibinfo  {journal} {Nature nanotechnology}\ }\textbf {\bibinfo
  {volume} {11}},\ \bibinfo {pages} {263} (\bibinfo {year} {2016})}\BibitemShut
  {NoStop}%
\bibitem [{\citenamefont {Lu}\ \emph {et~al.}(2012)\citenamefont {Lu},
  \citenamefont {Bark}, \citenamefont {Esque de~los Ojos}, \citenamefont
  {Alcala}, \citenamefont {Eom}, \citenamefont {Catalan},\ and\ \citenamefont
  {Gruverman}}]{Lu2012}%
  \BibitemOpen
  \bibfield  {author} {\bibinfo {author} {\bibfnamefont {H.}~\bibnamefont
  {Lu}}, \bibinfo {author} {\bibfnamefont {C.-W.}\ \bibnamefont {Bark}},
  \bibinfo {author} {\bibfnamefont {D.}~\bibnamefont {Esque de~los Ojos}},
  \bibinfo {author} {\bibfnamefont {J.}~\bibnamefont {Alcala}}, \bibinfo
  {author} {\bibfnamefont {C.~B.}\ \bibnamefont {Eom}}, \bibinfo {author}
  {\bibfnamefont {G.}~\bibnamefont {Catalan}}, \ and\ \bibinfo {author}
  {\bibfnamefont {A.}~\bibnamefont {Gruverman}},\ }\href {\doibase
  10.1126/science.1218693} {\bibfield  {journal} {\bibinfo  {journal}
  {Science}\ }\textbf {\bibinfo {volume} {336}},\ \bibinfo {pages} {59}
  (\bibinfo {year} {2012})}\BibitemShut {NoStop}%
\bibitem [{\citenamefont {Cross}(2006)}]{Cross06}%
  \BibitemOpen
  \bibfield  {author} {\bibinfo {author} {\bibfnamefont {L.}~\bibnamefont
  {Cross}},\ }\href@noop {} {\bibfield  {journal} {\bibinfo  {journal} {J.
  Mater. Sci.}\ }\textbf {\bibinfo {volume} {41}},\ \bibinfo {pages} {53}
  (\bibinfo {year} {2006})}\BibitemShut {NoStop}%
\bibitem [{\citenamefont {Martin}(1972)}]{Martin1972}%
  \BibitemOpen
  \bibfield  {author} {\bibinfo {author} {\bibfnamefont {R.~M.}\ \bibnamefont
  {Martin}},\ }\href {\doibase 10.1103/PhysRevB.5.1607} {\bibfield  {journal}
  {\bibinfo  {journal} {Phys. Rev. B}\ }\textbf {\bibinfo {volume} {5}},\
  \bibinfo {pages} {1607} (\bibinfo {year} {1972})}\BibitemShut {NoStop}%
\bibitem [{\citenamefont {Hamann}\ \emph {et~al.}(2005)\citenamefont {Hamann},
  \citenamefont {Wu}, \citenamefont {Rabe},\ and\ \citenamefont
  {Vanderbilt}}]{Hamann2005}%
  \BibitemOpen
  \bibfield  {author} {\bibinfo {author} {\bibfnamefont {D.~R.}\ \bibnamefont
  {Hamann}}, \bibinfo {author} {\bibfnamefont {X.}~\bibnamefont {Wu}}, \bibinfo
  {author} {\bibfnamefont {K.~M.}\ \bibnamefont {Rabe}}, \ and\ \bibinfo
  {author} {\bibfnamefont {D.}~\bibnamefont {Vanderbilt}},\ }\href {\doibase
  10.1103/PhysRevB.71.035117} {\bibfield  {journal} {\bibinfo  {journal} {Phys.
  Rev. B}\ }\textbf {\bibinfo {volume} {71}},\ \bibinfo {pages} {035117}
  (\bibinfo {year} {2005})}\BibitemShut {NoStop}%
\bibitem [{\citenamefont {Hong}\ and\ \citenamefont
  {Vanderbilt}(2013)}]{Hong2013}%
  \BibitemOpen
  \bibfield  {author} {\bibinfo {author} {\bibfnamefont {J.}~\bibnamefont
  {Hong}}\ and\ \bibinfo {author} {\bibfnamefont {D.}~\bibnamefont
  {Vanderbilt}},\ }\href {\doibase 10.1103/PhysRevB.88.174107} {\bibfield
  {journal} {\bibinfo  {journal} {Phys. Rev. B}\ }\textbf {\bibinfo {volume}
  {88}},\ \bibinfo {pages} {174107} (\bibinfo {year} {2013})}\BibitemShut
  {NoStop}%
\bibitem [{\citenamefont {Stengel}(2013{\natexlab{a}})}]{Stengel2013}%
  \BibitemOpen
  \bibfield  {author} {\bibinfo {author} {\bibfnamefont {M.}~\bibnamefont
  {Stengel}},\ }\href {\doibase 10.1103/PhysRevB.88.174106} {\bibfield
  {journal} {\bibinfo  {journal} {Phys. Rev. B}\ }\textbf {\bibinfo {volume}
  {88}},\ \bibinfo {pages} {174106} (\bibinfo {year}
  {2013}{\natexlab{a}})}\BibitemShut {NoStop}%
\bibitem [{\citenamefont {Stengel}(2014)}]{Stengel2014}%
  \BibitemOpen
  \bibfield  {author} {\bibinfo {author} {\bibfnamefont {M.}~\bibnamefont
  {Stengel}},\ }\href {\doibase 10.1103/PhysRevB.90.201112} {\bibfield
  {journal} {\bibinfo  {journal} {Phys. Rev. B}\ }\textbf {\bibinfo {volume}
  {90}},\ \bibinfo {pages} {201112} (\bibinfo {year} {2014})}\BibitemShut
  {NoStop}%
\bibitem [{\citenamefont {Dreyer}\ \emph {et~al.}(2018)\citenamefont {Dreyer},
  \citenamefont {Stengel},\ and\ \citenamefont {Vanderbilt}}]{Dreyer2018}%
  \BibitemOpen
  \bibfield  {author} {\bibinfo {author} {\bibfnamefont {C.~E.}\ \bibnamefont
  {Dreyer}}, \bibinfo {author} {\bibfnamefont {M.}~\bibnamefont {Stengel}}, \
  and\ \bibinfo {author} {\bibfnamefont {D.}~\bibnamefont {Vanderbilt}},\
  }\href {\doibase 10.1103/PhysRevB.98.075153} {\bibfield  {journal} {\bibinfo
  {journal} {Phys. Rev. B}\ }\textbf {\bibinfo {volume} {98}},\ \bibinfo
  {pages} {075153} (\bibinfo {year} {2018})}\BibitemShut {NoStop}%
\bibitem [{\citenamefont {Stengel}\ and\ \citenamefont
  {Vanderbilt}(2018)}]{quantum}%
  \BibitemOpen
  \bibfield  {author} {\bibinfo {author} {\bibfnamefont {M.}~\bibnamefont
  {Stengel}}\ and\ \bibinfo {author} {\bibfnamefont {D.}~\bibnamefont
  {Vanderbilt}},\ }\href {\doibase 10.1103/PhysRevB.98.125133} {\bibfield
  {journal} {\bibinfo  {journal} {Phys. Rev. B}\ }\textbf {\bibinfo {volume}
  {98}},\ \bibinfo {pages} {125133} (\bibinfo {year} {2018})}\BibitemShut
  {NoStop}%
\bibitem [{\citenamefont {Stengel}(2013{\natexlab{b}})}]{Stengel2013natcom}%
  \BibitemOpen
  \bibfield  {author} {\bibinfo {author} {\bibfnamefont {M.}~\bibnamefont
  {Stengel}},\ }\href {\doibase 10.1038/ncomms3693} {\bibfield  {journal}
  {\bibinfo  {journal} {Nat. Commun.}\ }\textbf {\bibinfo {volume} {4}},\
  \bibinfo {pages} {2693} (\bibinfo {year} {2013}{\natexlab{b}})}\BibitemShut
  {NoStop}%
\bibitem [{\citenamefont {Tagantsev}(1986)}]{Tagantsev-86}%
  \BibitemOpen
  \bibfield  {author} {\bibinfo {author} {\bibfnamefont {A.~K.}\ \bibnamefont
  {Tagantsev}},\ }\href {\doibase 10.1103/PhysRevB.34.5883} {\bibfield
  {journal} {\bibinfo  {journal} {Phys. Rev. B}\ }\textbf {\bibinfo {volume}
  {34}},\ \bibinfo {pages} {5883} (\bibinfo {year} {1986})}\BibitemShut
  {NoStop}%
\bibitem [{\citenamefont {Sakuri}\ and\ \citenamefont
  {Napolitano}(1994)}]{Sakuri}%
  \BibitemOpen
  \bibfield  {author} {\bibinfo {author} {\bibfnamefont {J.}~\bibnamefont
  {Sakuri}}\ and\ \bibinfo {author} {\bibfnamefont {J.}~\bibnamefont
  {Napolitano}},\ }\href@noop {} {\emph {\bibinfo {title} {Modern Quantum
  Mechanics--2nd ed.}}}\ (\bibinfo  {publisher} {Addison-Wesley, San Fransisco,
  CA},\ \bibinfo {year} {1994})\BibitemShut {NoStop}%
\bibitem [{\citenamefont {Tagantsev}\ and\ \citenamefont
  {Yurkov}(2012)}]{Tagantsev12}%
  \BibitemOpen
  \bibfield  {author} {\bibinfo {author} {\bibfnamefont {A.~K.}\ \bibnamefont
  {Tagantsev}}\ and\ \bibinfo {author} {\bibfnamefont {A.~S.}\ \bibnamefont
  {Yurkov}},\ }\href {\doibase https://doi.org/10.1063/1.4745037} {\bibfield
  {journal} {\bibinfo  {journal} {J. Appl. Phys.}\ }\textbf {\bibinfo {volume}
  {112}},\ \bibinfo {pages} {044103} (\bibinfo {year} {2012})}\BibitemShut
  {NoStop}%
\bibitem [{\citenamefont {Kleinman}\ and\ \citenamefont
  {Bylander}(1982)}]{Kleinman1982}%
  \BibitemOpen
  \bibfield  {author} {\bibinfo {author} {\bibfnamefont {L.}~\bibnamefont
  {Kleinman}}\ and\ \bibinfo {author} {\bibfnamefont {D.~M.}\ \bibnamefont
  {Bylander}},\ }\href {\doibase 10.1103/PhysRevLett.48.1425} {\bibfield
  {journal} {\bibinfo  {journal} {Phys. Rev. Lett.}\ }\textbf {\bibinfo
  {volume} {48}},\ \bibinfo {pages} {1425} (\bibinfo {year}
  {1982})}\BibitemShut {NoStop}%
\bibitem [{\citenamefont {Ismail-Beigi}\ \emph {et~al.}(2001)\citenamefont
  {Ismail-Beigi}, \citenamefont {Chang},\ and\ \citenamefont
  {Louie}}]{ICL2001}%
  \BibitemOpen
  \bibfield  {author} {\bibinfo {author} {\bibfnamefont {S.}~\bibnamefont
  {Ismail-Beigi}}, \bibinfo {author} {\bibfnamefont {E.~K.}\ \bibnamefont
  {Chang}}, \ and\ \bibinfo {author} {\bibfnamefont {S.~G.}\ \bibnamefont
  {Louie}},\ }\href {\doibase 10.1103/PhysRevLett.87.087402} {\bibfield
  {journal} {\bibinfo  {journal} {Phys. Rev. Lett.}\ }\textbf {\bibinfo
  {volume} {87}},\ \bibinfo {pages} {087402} (\bibinfo {year}
  {2001})}\BibitemShut {NoStop}%
\bibitem [{\citenamefont {Essin}\ \emph {et~al.}(2010)\citenamefont {Essin},
  \citenamefont {Turner}, \citenamefont {Moore},\ and\ \citenamefont
  {Vanderbilt}}]{Essin2010}%
  \BibitemOpen
  \bibfield  {author} {\bibinfo {author} {\bibfnamefont {A.~M.}\ \bibnamefont
  {Essin}}, \bibinfo {author} {\bibfnamefont {A.~M.}\ \bibnamefont {Turner}},
  \bibinfo {author} {\bibfnamefont {J.~E.}\ \bibnamefont {Moore}}, \ and\
  \bibinfo {author} {\bibfnamefont {D.}~\bibnamefont {Vanderbilt}},\ }\href
  {\doibase 10.1103/PhysRevB.81.205104} {\bibfield  {journal} {\bibinfo
  {journal} {Phys. Rev. B}\ }\textbf {\bibinfo {volume} {81}},\ \bibinfo
  {pages} {205104} (\bibinfo {year} {2010})}\BibitemShut {NoStop}%
\bibitem [{\citenamefont {Gonze}(1997)}]{Gonze1997}%
  \BibitemOpen
  \bibfield  {author} {\bibinfo {author} {\bibfnamefont {X.}~\bibnamefont
  {Gonze}},\ }\href {\doibase 10.1103/PhysRevB.55.10337} {\bibfield  {journal}
  {\bibinfo  {journal} {Phys. Rev. B}\ }\textbf {\bibinfo {volume} {55}},\
  \bibinfo {pages} {10337} (\bibinfo {year} {1997})}\BibitemShut {NoStop}%
\bibitem [{\citenamefont {Stengel}\ and\ \citenamefont
  {Vanderbilt}(2016)}]{StengelChapter}%
  \BibitemOpen
  \bibfield  {author} {\bibinfo {author} {\bibfnamefont {M.}~\bibnamefont
  {Stengel}}\ and\ \bibinfo {author} {\bibfnamefont {D.}~\bibnamefont
  {Vanderbilt}},\ }in\ \href@noop {} {\emph {\bibinfo {booktitle}
  {Flexoelectricity in Solids From Theory to Applications}}},\ \bibinfo
  {editor} {edited by\ \bibinfo {editor} {\bibfnamefont {A.~K.}\ \bibnamefont
  {Tagantsev}}\ and\ \bibinfo {editor} {\bibfnamefont {P.~V.}\ \bibnamefont
  {Yudin}}}\ (\bibinfo  {publisher} {World Scientific Publishing Co.},\
  \bibinfo {address} {Singapore},\ \bibinfo {year} {2016})\ Chap.~\bibinfo
  {chapter} {2}, pp.\ \bibinfo {pages} {31--110}\BibitemShut {NoStop}%
\bibitem [{\citenamefont {Gonze}\ and\ \citenamefont
  {Lee}(1997)}]{Abinit_phonon_2}%
  \BibitemOpen
  \bibfield  {author} {\bibinfo {author} {\bibfnamefont {X.}~\bibnamefont
  {Gonze}}\ and\ \bibinfo {author} {\bibfnamefont {C.}~\bibnamefont {Lee}},\
  }\href {\doibase 10.1103/PhysRevB.55.10355} {\bibfield  {journal} {\bibinfo
  {journal} {Phys. Rev. B}\ }\textbf {\bibinfo {volume} {55}},\ \bibinfo
  {pages} {10355} (\bibinfo {year} {1997})}\BibitemShut {NoStop}%
\bibitem [{\citenamefont {Perdew}\ and\ \citenamefont {Wang}(1992)}]{Perdew92}%
  \BibitemOpen
  \bibfield  {author} {\bibinfo {author} {\bibfnamefont {J.~P.}\ \bibnamefont
  {Perdew}}\ and\ \bibinfo {author} {\bibfnamefont {Y.}~\bibnamefont {Wang}},\
  }\href {\doibase 10.1103/PhysRevB.45.13244} {\bibfield  {journal} {\bibinfo
  {journal} {Phys. Rev. B}\ }\textbf {\bibinfo {volume} {45}},\ \bibinfo
  {pages} {13244} (\bibinfo {year} {1992})}\BibitemShut {NoStop}%
\bibitem [{\citenamefont {Troullier}\ and\ \citenamefont
  {Martins}(1991)}]{Troullier91}%
  \BibitemOpen
  \bibfield  {author} {\bibinfo {author} {\bibfnamefont {N.}~\bibnamefont
  {Troullier}}\ and\ \bibinfo {author} {\bibfnamefont {J.~L.}\ \bibnamefont
  {Martins}},\ }\href {\doibase 10.1103/PhysRevB.43.1993} {\bibfield  {journal}
  {\bibinfo  {journal} {Phys. Rev. B}\ }\textbf {\bibinfo {volume} {43}},\
  \bibinfo {pages} {1993} (\bibinfo {year} {1991})}\BibitemShut {NoStop}%
\bibitem [{\citenamefont {Fuchs}\ and\ \citenamefont
  {Scheffler}(1999)}]{fhi98PP}%
  \BibitemOpen
  \bibfield  {author} {\bibinfo {author} {\bibfnamefont {M.}~\bibnamefont
  {Fuchs}}\ and\ \bibinfo {author} {\bibfnamefont {M.}~\bibnamefont
  {Scheffler}},\ }\href@noop {} {\bibfield  {journal} {\bibinfo  {journal}
  {Comput. Phys. Commun.}\ }\textbf {\bibinfo {volume} {119}},\ \bibinfo
  {pages} {67} (\bibinfo {year} {1999})}\BibitemShut {NoStop}%
\bibitem [{\citenamefont {Monkhorst}\ and\ \citenamefont
  {Pack}(1976)}]{Monkhorst76}%
  \BibitemOpen
  \bibfield  {author} {\bibinfo {author} {\bibfnamefont {H.~J.}\ \bibnamefont
  {Monkhorst}}\ and\ \bibinfo {author} {\bibfnamefont {J.~D.}\ \bibnamefont
  {Pack}},\ }\href@noop {} {\bibfield  {journal} {\bibinfo  {journal} {Comput.
  Phys. Commun.}\ }\textbf {\bibinfo {volume} {13}},\ \bibinfo {pages} {5188}
  (\bibinfo {year} {1976})}\BibitemShut {NoStop}%
\bibitem [{\citenamefont {Portigal}\ and\ \citenamefont
  {Burstein}(1968)}]{Portigal1968}%
  \BibitemOpen
  \bibfield  {author} {\bibinfo {author} {\bibfnamefont {D.~L.}\ \bibnamefont
  {Portigal}}\ and\ \bibinfo {author} {\bibfnamefont {E.}~\bibnamefont
  {Burstein}},\ }\href {\doibase 10.1103/PhysRev.170.673} {\bibfield  {journal}
  {\bibinfo  {journal} {Phys. Rev.}\ }\textbf {\bibinfo {volume} {170}},\
  \bibinfo {pages} {673} (\bibinfo {year} {1968})}\BibitemShut {NoStop}%
\bibitem [{\citenamefont {Stengel}(2016)}]{Stengel2016}%
  \BibitemOpen
  \bibfield  {author} {\bibinfo {author} {\bibfnamefont {M.}~\bibnamefont
  {Stengel}},\ }\href {\doibase 10.1103/PhysRevB.93.245107} {\bibfield
  {journal} {\bibinfo  {journal} {Phys. Rev. B}\ }\textbf {\bibinfo {volume}
  {93}},\ \bibinfo {pages} {245107} (\bibinfo {year} {2016})}\BibitemShut
  {NoStop}%
\end{thebibliography}%

\end{document}